\documentclass[preprint,12pt,numbers,sort&compress,square]{elsarticle}

\usepackage[english]{babel}

\usepackage[letterpaper,top=2cm,bottom=2cm,left=3cm,right=3cm,marginparwidth=1.75cm]{geometry}
\usepackage{amsmath}
\usepackage{cancel}
\usepackage{braket}
\usepackage{graphicx}
\usepackage{xcolor}
\usepackage[export]{adjustbox}
\usepackage[colorlinks=true, allcolors=blue]{hyperref}
\usepackage{caption}
\usepackage{subcaption}
\usepackage{multirow}
\usepackage{lineno}
\usepackage{color,soul}
\usepackage{nomencl}
\usepackage[normalem]{ulem}
\usepackage{sectsty}
\sectionfont{\normalfont\bfseries} 
\subsectionfont{\normalfont\mdseries} 
\subsubsectionfont{\itshape\mdseries} 
\usepackage{booktabs} 
\usepackage{float}
\usepackage{textgreek}

\makenomenclature

\begin{document}

\begin{frontmatter}

\title{An integrated viscoelastic modeling framework combining analytical 
and FEM approaches: application to WSe\textsubscript{2} coatings}

\author[add1]{Mohamed Bensalem\corref{cor1}}
\ead{mohamed.bensalem@cvut.cz}

\author[add1]{Fateh Bahadur}
\ead{bahadfat@fel.cvut.cz}

\author[add1]{Yue Wang}
\ead{wangyue@cvut.cz}

\author[add1]{Nabil Daghbouj}
\ead{daghbnab@fel.cvut.cz}

\author[add1,add2]{Tomas Polcar\corref{cor1}}
\ead{polcatom@fel.cvut.cz}

\cortext[cor1]{Corresponding authors: Mohamed Bensalem (mohamed.bensalem@cvut.cz), Tomas Polcar (polcatom@fel.cvut.cz)}

\affiliation[add1]{
  organization={Department of Control Engineering, Czech Technical University in Prague},
  addressline={},
  city={Prague},
  postcode={121 35},
  country={Czech Republic}
}

\affiliation[add2]{
  organization={School of Engineering, University of Southampton},
  addressline={Highfield},
  city={Southampton},
  postcode={SO17 1BJ},
  country={United Kingdom}
}

\begin{abstract}

This work presents an integrated methodology combining analytical and finite element (FEM)-based viscoelastic modeling for characterizing the nanoindentation response of coatings. The proposed framework combines two complementary modeling approaches: an analytical model based on the Burgers formulation to analyze nanoindentation load-displacement data and extract rheological parameters, and a numerical FEM model implemented in ABAQUS using a 2D axisymmetric indenter–coating-substrate configuration with viscoelasticity represented through Prony series. An automated inverse optimization routine employing the Nelder–Mead simplex algorithm minimizes the discrepancy between experimental and simulated responses. The methodology is demonstrated and validated on WSe\textsubscript{2} coatings, showing an agreement between the analytical predictions, FEM simulations, and experimental measurements. Despite the geometric simplifications, the FEM approach provides accurate predictions while maintaining high computational efficiency of time-dependent mechanical behavior. The proposed framework provides a robust tool for characterizing viscoelastic behavior from nanoindentation data while enabling access to internal stress and strain fields, thereby offering deeper insight into plastic deformation, crack initiation, and failure mechanism.

\end{abstract}

\begin{graphicalabstract}
\includegraphics[width=1\linewidth]{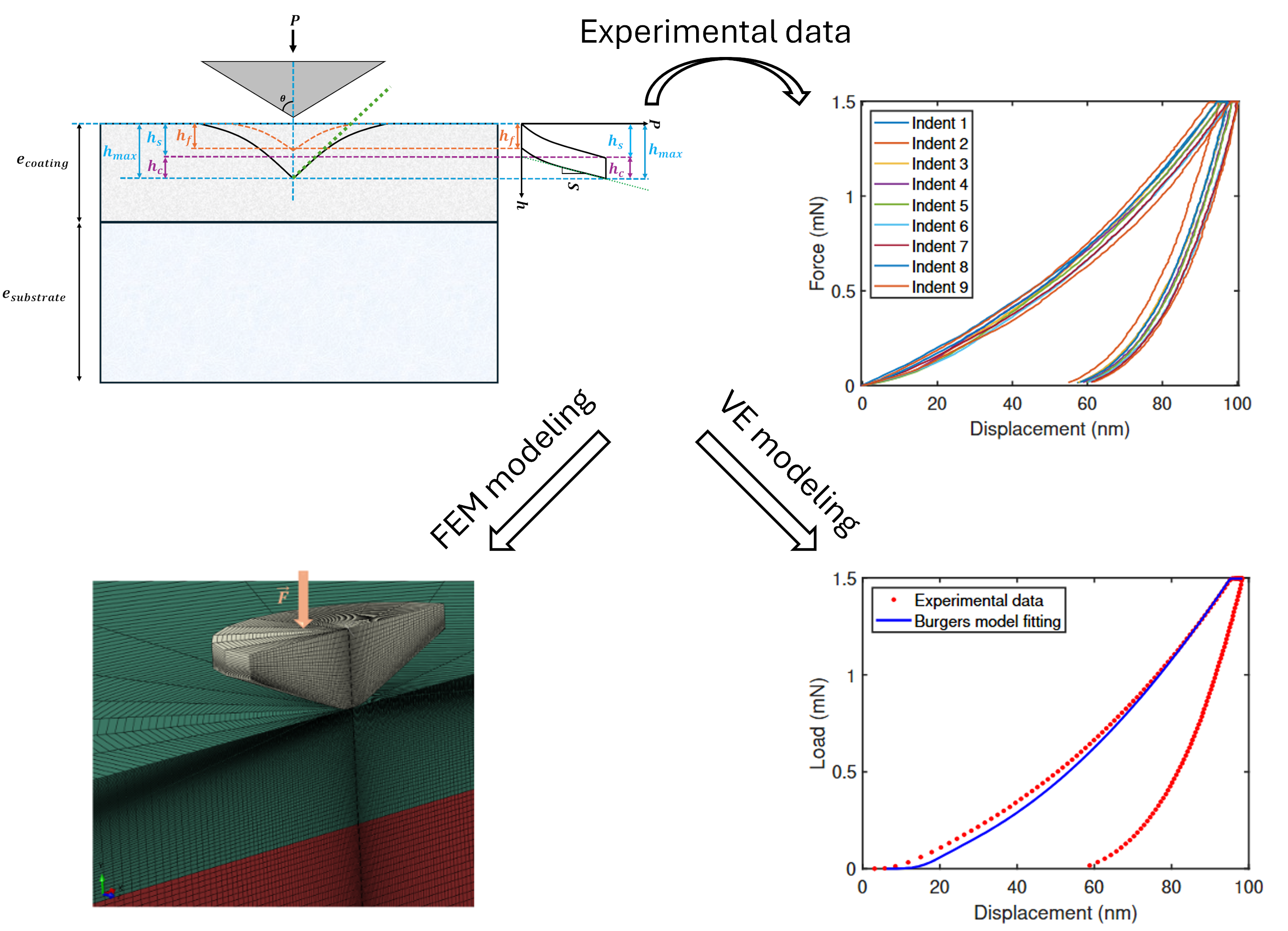}
\end{graphicalabstract}

\begin{highlights}
\item Developed an integrated methodology combining analytical Burgers viscoelastic model and FEM-based simulations for nanoindentation 
characterization.
\item Implemented a 2D axisymmetric FEM model in ABAQUS with viscoelasticity represented through Prony series for computational efficiency.
\item Applied an automated inverse optimization routine using the Nelder-Mead algorithm to optimize viscoelastic parameters.
\item Validated the framework on WSe\textsubscript{2} coatings, achieving good agreement between experimental and modeled load-displacement behavior.
\end{highlights}


\end{frontmatter}

\section{Introduction}\label{Introdduction}

Understanding the mechanical behavior of thin coatings is critical for designing robust layered systems that operate under extreme conditions such as severe mechanical contact, high temperatures, and irradiation. Accurate characterization methods enable the prediction of stress–strain fields within coatings and substrates, which is essential for improving performance in demanding environments. While numerous coating materials have been developed to enhance tribological and mechanical properties \cite{110,111}, the reliability of these improvements strongly depends on the ability to characterize their mechanical response at small scales.

Nanoindentation has emerged as a widely adopted technique for estimating mechanical properties such as hardness and elastic modulus from load–
displacement data \cite{15,16}. Its suitability for thin films and tribological coatings stems from its ability to probe very small material volumes. While nanoindentation involves both elastic and plastic deformation during loading, unloading begins with elastic recovery. Therefore, the contact stiffness at the initial stage of unloading is used to extract mechanical properties through analytical approaches such as the Oliver-Pharr method \cite{105}. Based on Sneddon's solution for the indentation of an elastic half-space by a rigid axisymmetric indenter \cite{15,16, 106}, the Oliver–Pharr method enables the determination of Young’s modulus and hardness from the contact area at maximum load. However, this classical approach assumes purely elastic–plastic behavior and does not account for time-dependent deformation, which may be observed in polymers \cite{107,108} and in some thin-film and coating systems during the hold phase of indentation tests \cite{109}.

Viscoelastic modeling has been widely employed to describe the behavior of time-dependent materials using analytical \cite{100,101,102}, semi-analytical \cite{103} and numerical approaches \cite{100}. Analytical formulations commonly rely on combination of spring and dashpot elements to represent the stress-strain and stress-strain rate relationships, respectively \cite{100,104}. Depending on the complexity of the material response, several rheological models have been proposed, including Burgers \cite{v1,v2,25,v4} and generalized Maxwell representations \cite{100,104}. Such models have also been used to interpret nanoindentation creep data and to compensate for time-dependent deformation effects that are not captured by the classical Oliver–Pharr analysis \cite{v2,v6}. In parallel, finite element modeling (FEM) has become an effective tool for simulating viscoelastic behavior, enabling visualization of stress and strain fields while facilitating the identification of creep and relaxation parameters \cite{v3}. More recently, FEM approaches have been integrated into elastoplastic behavior to evaluate the toughness of hard coatings such as W-C coatings \cite{17,18,19} and TiN coatings \cite{20} where a pop-in events in the load-unload curve were used as an indicator of crack initiation. A similar methodology was applied to TiAlN coatings \cite{21}.

Despite these developments, studies directly comparing analytical viscoelastic formulations and inverse FEM-based identification approaches for the characterization of coating materials remain limited. Furthermore, the respective advantages, limitations, and consistency of both approaches when applied to the same experimental nanoindentation data are not fully understood.

In this work, a comprehensive methodology for the mechanical characterization of thin coatings under nanoindentation is proposed by employing two independent approaches. The first approach is based on an analytical Burgers viscoelastic model, while the second relies on simplified 2D-axisymmetric finite element simulations coupled with an automated Nelder–Mead inverse optimization procedure implemented within ABAQUS. In the numerical framework, the experimental load–displacement response is used as a reference and the Young's modulus together with the Prony-series coefficients are iteratively updated. At each iteration, the error between the experimental and the simulated displacement is computed and minimized using minimize package within PYTHON until convergence is achieved. The results obtained from both approaches are subsequently compared and discussed to assess their respective capabilities for describing the viscoelastic behavior of the coating.

To demonstrate the applicability of the proposed methodology, tungsten diselenide (WSe\textsubscript{2}) coatings deposited via magnetron sputtering were selected as a case study. WSe\textsubscript{2} belongs to the family of layered transition metal dichalcogenides (TMDs), which are widely recognized for their excellent tribological performance under severe mechanical contact, elevated temperatures, and irradiation environments \cite{1}. Previous studies have shown that WSe\textsubscript{2} exhibits superior tribological performance at temperatures up to 400 °C and in humid atmosphere, outperforming MoS\textsubscript{2} in terms of friction stability \cite{13,14}. Despite the extensive literature devoted to its tribological properties, its time-dependent mechanical response under nanoindentation remains insufficiently explored, particularly from a viscoelastic perspective.

By applying both the analytical Burgers model and the FEM-based inverse identification approach to the same experimental nanoindentation data, the present work evaluates their respective capabilities for describing the viscoelastic behavior of WSe\textsubscript{2} coatings. In addition to providing a detailed characterization of the coating's time-dependent deformation response, the study highlights the advantages and limitations of each methodology and assesses their consistency in estimating viscoelastic material parameters.

\section{Nanoindentation} \label{Nanoindentation}

Hysitron TI 950 TriboIndenter is used for the indentation experiments. This device offers high precision and control over the indentation process, making it suitable for the nanoscale characterization. It uses a diamond indenter and is equipped with a sophisticated closed-loop force control system that enables the application of highly controlled loads and precise measurement of indentation depths. It supports a wide range of loads, from micronewtons to millinewtons, which is crucial for accurately assessing the hardness and modulus of materials with varying mechanical properties. The system’s high displacement resolution, down to the sub-nanometer level, allows for detailed analysis of material deformation. TriboIndenter TI 950 offers multiple indentation modes, including quasi static mode, partial unloading mode and dynamic mechanical analysis, providing a comprehensive evaluation of the material’s mechanical behavior. It also includes environmental control features to maintain consistent testing conditions, which is vital for accurate and reproducible results.

Indentation tests were performed at maximum load of 1.5 mN with sufficient distance between each indentation (30 \textmu m) to ensure accurate measurements. The used indenter was a Berkovich probe made of standard diamond (E\textsubscript{indenter} = 1140 GPa, \textnu = 0.07). The nanoindentation experiments have been conducted in ambient temperature on coated steel with thickness of 1 \textmu m of WSe\textsubscript{2} on 5 mm depth of substrate. The applied indentation load is set in a way that the indentation depth to be 10\%-15\% of the total thickness of the coating  that is 1 \textmu m allowing to avoid the contribution of the substrate in the characterization. Thermal drift has been corrected using a preset of 0.05 nm/s. However, in this study, we focus solely on studying a single indentation depth in order to validate the method of 
characterization which will be used in future for more different coatings. The experimental data will be coupled with FEM model to investigate 
the mechanical properties of the studied coating.

\begin{figure}[t]
    \centering
    \begin{subfigure}[t]{0.5\textwidth}
        \centering
        \adjustbox{valign=t}{\includegraphics[width=\linewidth]{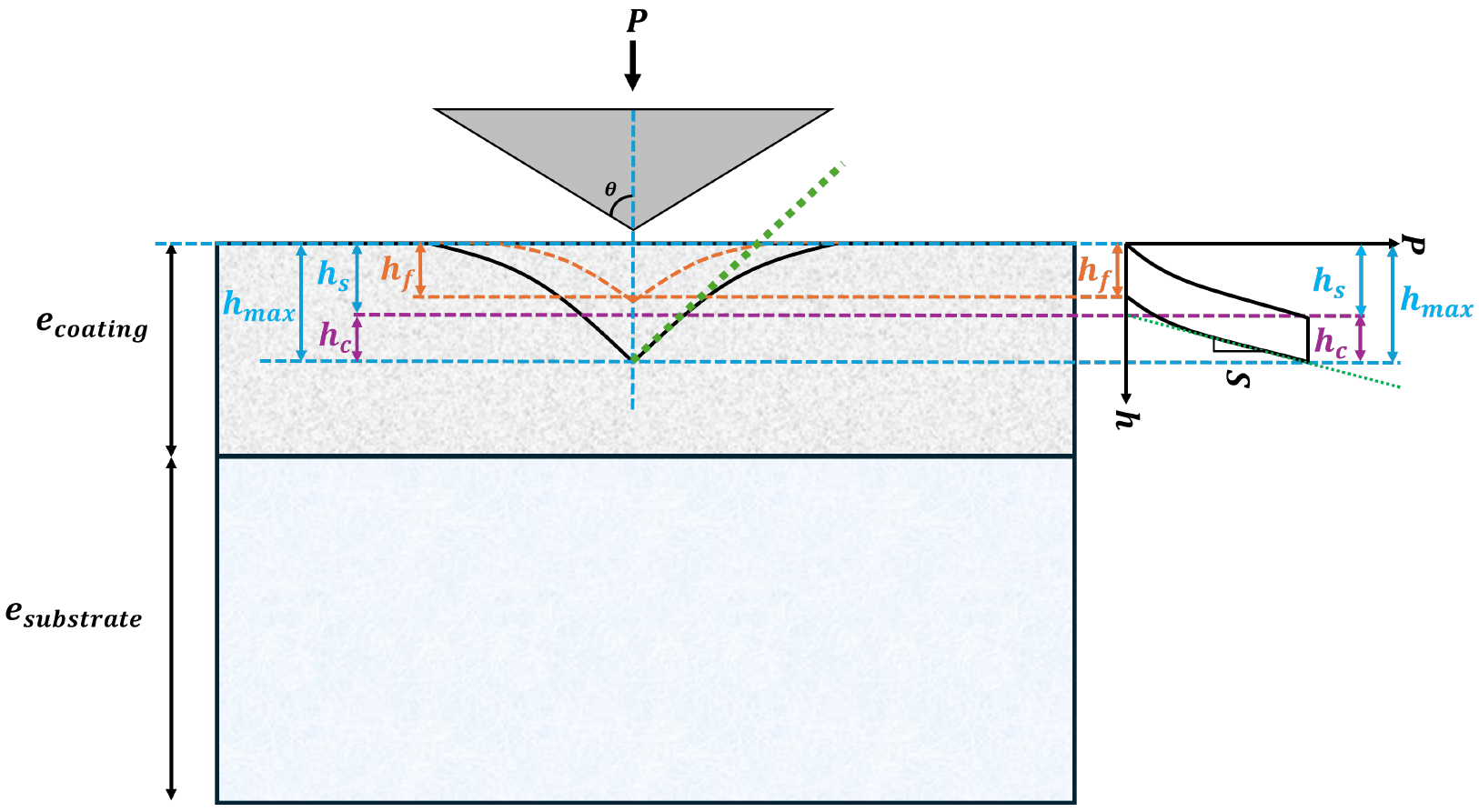}}
        \vspace{3.6em} 
        \caption{}
        \label{fig1a}
    \end{subfigure}
    \hfill
    \begin{subfigure}[t]{0.47\textwidth}
        \centering
        \adjustbox{valign=t}{\includegraphics[width=\linewidth]{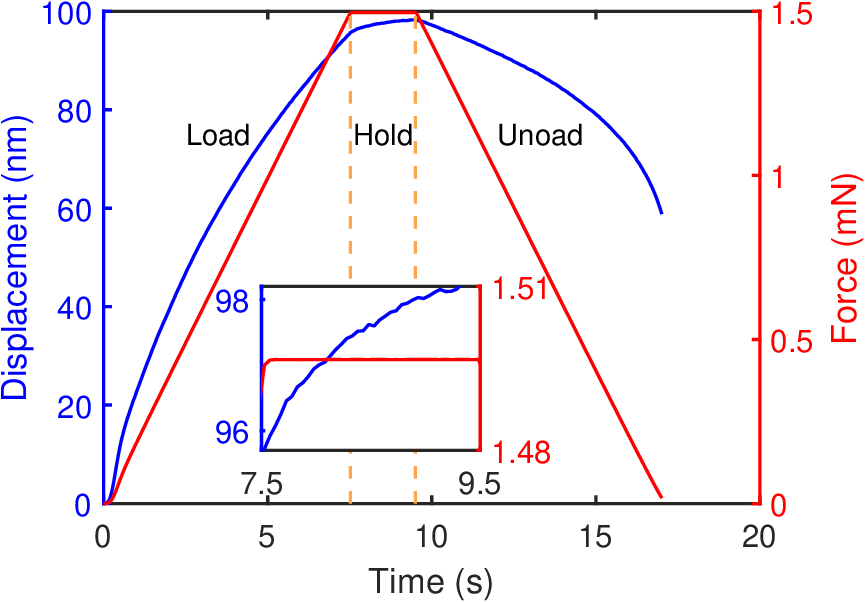}}
        \vspace{0.5em} 
        \caption{}
        \label{figa1b}
    \end{subfigure}

    \caption{(a) Schematic of the indentation using a Berkovich-equivalent conical indenter ($\theta = 70.3^\circ$); (b) Experimental 
    nanoindentation performed at a maximum load of 1.5 mN and a strain rate of 0.05 s$^{-1}$, showing a hold phase that indicates viscoelastic 
    behavior.}
    \label{fig1}
\end{figure}

Nanoindentation trapezoidal loading profile represents three segments; loading phase starting from 0 mN up to first inflection point at 1.5 mN 
during 7.5 s, holding phase keep the loading at 1.5 mN for 2 s, and unloading phase occurs after the second inflection point form 1.5 mN to 0 
mN during 7.5 s. The applied load was continuously adjusted in the load phase to compensate for the displacement of the indenter support 
springs. This dynamic compensation ensured that the load exerted on the sample remained constant throughout the entire holding period, which is 
critical for accurate creep characterization. Prior to conducting the measurements, a standard sapphire Al\textsubscript{2}O\textsubscript{3} 
sample was used for calibration to verify the accuracy and reliability of the testing system. The illustration of nanoindentation, the imposed 
load and the resulting displacement are shown in \hyperref[fig1]{Fig.~\ref{fig1}}.

\section{Constitutive model} \label{Constitutive model}

We consider that the indentation is carried out using rigid cone of half-angle   $\theta$ in linear viscoelastic material, where the load $F(t)$ is 
controlled function of time using maximum indentation force $F_\mathrm{max}$ and history function $F(t)$, for example in case of linear load, 
hold and linear unload, the load form is given by equation \hyperref[eq1]{eq.~\ref{eq1}}.

\begin{equation}
F(t) = 
\begin{cases}
\displaystyle \frac{F_{\text{max}}}{t_1} t & \text{for } 0 \leq t \leq t_1 \\[8pt]
F_{\text{max}} & \text{for } t_1 < t \leq t_2 \\[8pt]
\displaystyle \frac{F_{\text{max}}}{t_3 - t_2} (t_3 - t) & \text{for } t_2 < t \leq t_3
\end{cases}
\label{eq1}
\end{equation}

\begin{figure}[t]
    \centering
    \includegraphics[width=0.7\linewidth]{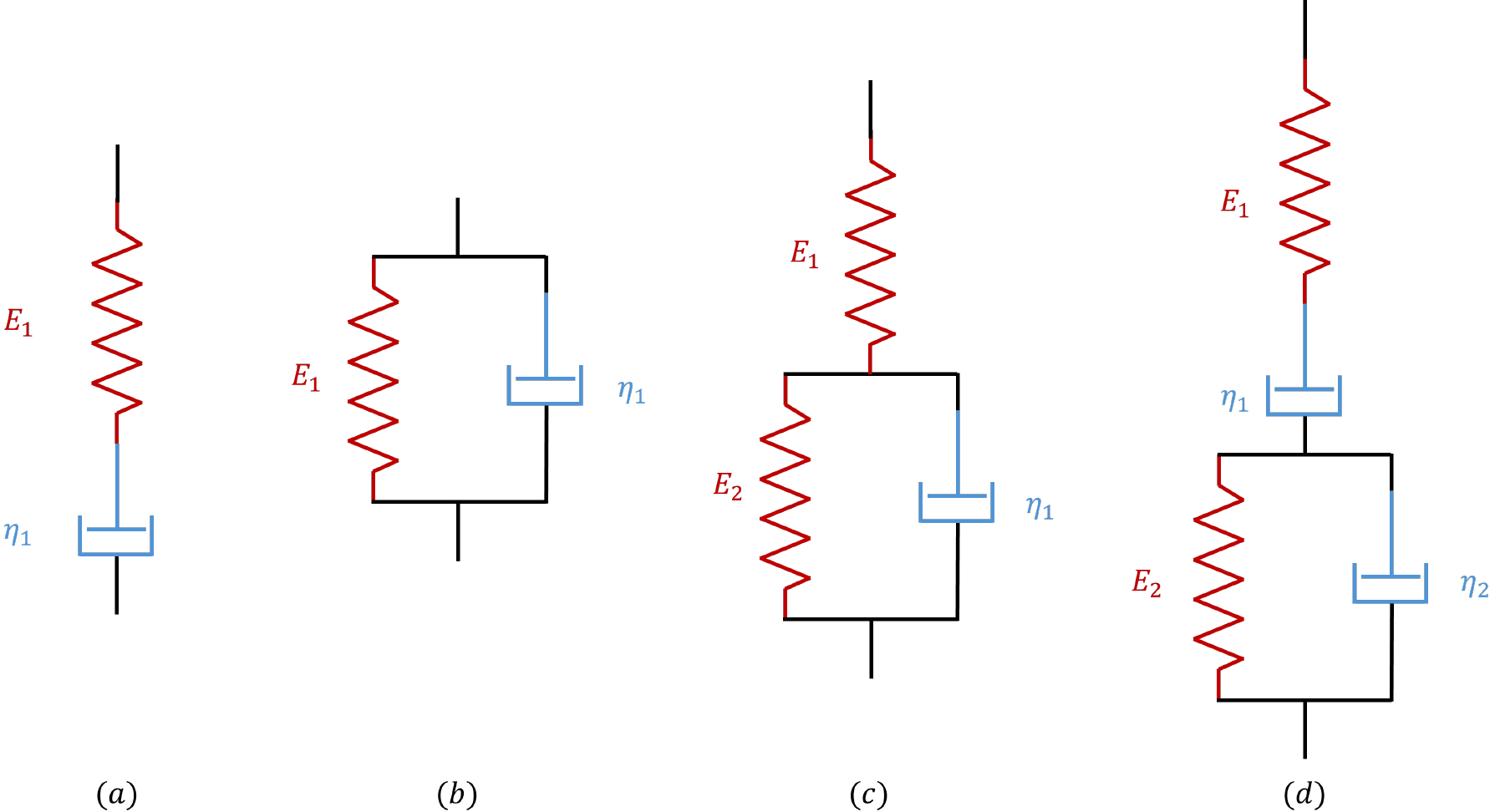}
    \caption{Viscoelastic behavior modeling using : (a) Maxwell, (b) Kelvin-Voigt, (c) Maxwell-Kelvin-Voigt and (d) Burgers model.}
    \label{fig2}
\end{figure}

The constitutive model of viscoelastic behavior uses basically the spring and the dashpot as basic elements to describe the elastic and viscoelastic contributions. The spring is represented by the elastic modulus $E$ and the dashpot is represented by the viscosity $\eta$. These elements can be connected in series constituting the Maxwell model, in parallel constituting Kelvin-Voigt model or combined in series and in parallel which is known as Burgers model. The number of elements starts from 2 to infinite depending on the viscoelastic response of the material following to an excitation. The most common schemes of modeling are given in \hyperref[fig2]{Fig.~\ref{fig2}}.

For any linear isotropic viscoelastic material, the general constitutive relationship between the deviatoric stress $S_{ij}$ and 
strain $d_{ij}$, and between the volumetric stress $\sigma_{ij}$ and strain $\epsilon_{ij}$ \cite{24}, can be given in the case of 
rigid and frictionless conical indenter by \hyperref[eq2]{eq.~\ref{eq2}}.

\begin{equation}
\begin{cases}
S_{ij}(t)      = \displaystyle2\int_0^{t} G(t-\tau) \frac{\partial d_{ij}(\tau)}{\partial\tau} d\tau \\[0.8em]
\sigma_{ii}(t) = \displaystyle3\int_0^{t} K(t - \tau) \frac{d\epsilon_{ii}(\tau)}{d\tau} d\tau
\end{cases}
\label{eq2}
\end{equation}

Where $G(t)$ and $K(t)$ are the relaxation moduli in shear and dilatation respectively. The \hyperref[eq2]{eq.~\ref{eq2}} has another 
inverted form as given by \hyperref[eq3]{eq.~\ref{eq3}} 

\begin{equation}
\begin{cases}
2\,d_{ij}(t)       = \displaystyle\int_0^{t} J_1(t-\tau)\,\frac{\partial S_{ij}(\tau)}{\partial\tau}\,d\tau \\[0.8em]
3\,\epsilon_{ii}(t) = \displaystyle\int_0^{t} J_2(t-\tau)\,\frac{d\sigma_{ii}(\tau)}{d\tau}\,d\tau
\end{cases}
\label{eq3}
\end{equation}

Where $J_1(t)$ and $J_2(t)$ are shear and volumetric compliance respectively.

In this work, the viscoelastic response is modeled primarily through deviatoric creep rather than volumetric creep since the indentation in linear viscoelastic solids is dominated by shear deformation. By assuming that a time-independent Poisson’s ratio, the material behavior under indentation can be described by the shear relaxation function $G(t)$ and Poisson’s ratio $\nu$. Since conical and pyramidal indentation generates stress fields that are largely deviatoric, volumetric changes are negligible. Therefore, focusing on shear compliance or relaxation functions simplifies modeling without loss of accuracy for typical viscoelastic solids \cite{22}.

The equation describing the Maxwell and Kelvin-Voigt (KV) is formulated as:

\begin{equation}
\begin{cases}
\displaystyle2\,\dot{\epsilon}^d(t)_{\text{Maxwell}} = \frac{\dot{\sigma}^d(t)}{G_1} + \frac{\sigma^d(t)}{\eta_1} \\[0.8em]
\displaystyle2\,\dot{\epsilon}^d(t)_{\text{KV}} + \frac{G_2}{\eta_2}\,2\,\dot{\epsilon}^d(t)_{KV} = \frac{\sigma^d(t)}{\eta_2}
\end{cases}
\label{eq4}
\end{equation}

Therefore, the solution of the both system is expressed as:

\begin{equation}
\begin{cases}
\displaystyle2\,\epsilon^d(t)_{\text{Maxwell}} = \sigma^d(t) \otimes \left[ \frac{1}{G_1} + \frac{t}{\eta_1} \right] \\[0.8em]
\displaystyle2\,\epsilon^d(t)_{\text{KV}} = \sigma^d(t) \otimes \left[\frac{1}{G_2}\left( 1 - e^{-\frac{G_2t}{\eta_2}} \right) 
\right] \\
\end{cases}
\label{eq5}
\end{equation}

Then, the Burgers model that will be chosen in this work to describe the viscoelastic behavior can be given by total strain of the two blocks 
Maxwell and KV in series as given by \hyperref[eq6]{eq.~\ref{eq6}}.

\begin{equation}
\displaystyle2\,\epsilon^d(t)_{KV} = \sigma^d(t) \otimes \underbrace{\left[\frac{1}{G_1} + \frac{t}{\eta_1} + \frac{1}{G_2}\left( 1 - e^{-
\frac{G_2t}{\eta_2}} \right) \right]}_{J_1(t)}
\label{eq6}
\end{equation}

The compliance function ${J_1(t)}$ represents the tensor of creep of the material. The compliance function is used in the load-controlled 
testing. However, the relaxation function ${G(t)}$ is used in the case of displacement-controlled testing. The governing equation for conical 
indenter in elastic homogeneous viscoelastic medium must include the load history, elastic properties, viscous properties and indenter geometry 
factor, it given by Galin-Sneddon solution \cite{22,24} as shown by \hyperref[eq7]{eq.~\ref{eq7}}.

\begin{equation}
h^2(t) = \frac{\pi}{2\tan\theta} \frac{1-\nu^2}{E} F(t) = \frac{\pi}{2\tan\theta} \frac{F(t)}{E_r} = \frac{\pi\,(1-
\nu)}{4\tan\theta} \frac{F(t)}{G}
\label{eq7}
\end{equation}

Where $E$, $E_r$, and $\nu$ are the Young's modulus, the indentation modulus (or reduced modulus in case of rigid indenter 
$\cancelto{negligible}{\frac{1-\nu_{ind}^2}{E_{ind}})}$, and the Poisson's ratio, with $\left ( E_r = \frac{E}{1 - \nu^2} \right)$ and 
the shear modulus $G$ is given by $\left( G = \frac{E}{2(1+\nu)} \right)$. The \hyperref[eq7]{eq.~\ref{eq7}} is valid if the force is 
independent variable, but when the force is time-dependent, the general form allowing obtaining the indentation displacement is given by the 
\hyperref[eq8]{eq.~\ref{eq8}} \cite{24,25}.

\begin{equation}
\displaystyle h^2(t) = \underbrace{\frac{\pi (1-\nu)}{4 \tan\theta}}_{Constant} \int_0^{t} J_1(t - \tau) \frac{dF(\tau)}
{d\tau} d\tau
\label{eq8}
\end{equation}

\noindent\textbf{$\bullet$ Ramp phase ($0 \leq t \leq t_1$)}
During the ramp, $F(t) = \frac{F_{max}}{t_1} t$ (\hyperref[eq1]{eq.~\ref{eq1}}), so $\frac{dF(t)}{dt} dt = \frac{F_{max}}{t_1}$ 
the equation (\hyperref[eq8]{eq.~\ref{eq8}}) becomes:

\begin{equation}
h^2(t) = \frac{\pi (1-\nu)}{4 \tan\theta} \frac{F_{max}}{t_1} \int_0^{t} J_1(t - \tau) d\tau
\label{eq9}
\end{equation}

By using the variable change $u=t-\tau$, thus, $du = -d\tau$ and if $\tau=0, u=t$ and if $\tau=t, u=0$, Therefore the \hyperref[eq9]
{eq.~\ref{eq9}} can be rewritten:

\begin{equation}
h^2(t) = \frac{\pi (1-\nu)}{4 \tan\theta} \frac{F_{max}}{t_1} \left( \frac{t}{G_1} + 
\frac{t^2}{2\eta_1} + \frac{1}{G_2} \left(t - \tau_2 + \tau_2e^{- \frac{t}{\tau_2}}\right)\right), \quad \text{with } \tau_2 = 
\frac{\eta_2}{G_2}
\label{eq10}
\end{equation}

\noindent\textbf{$\bullet$ Hold phase ($t_1 < t \leq t_2$)}
The hold phase response represents the accumulation of response history from $t=0$ until the instant $t_1<t\leq t_2$, therefore, the 
(\hyperref[eq9]{eq.~\ref{eq9}}) becomes:

\begin{equation}
h^2(t) = \frac{\pi (1-\nu)}{4 \tan\theta} \left( \int_0^{t_1} J_1(t - \tau) \frac{dF(\tau)}{d\tau} d\tau + \int_{t_1}^{t} 
J_1(t - \tau) \cancelto{0}{\frac{dF(\tau)}{d\tau}} d\tau \right)
\label{eq11}
\end{equation}

Similarly, the solution for $t_1<t\leq t_2$ obtained by resolving the following equation given by:
 
\begin{equation}
h^2(t) = \frac{\pi (1-\nu)}{4 \tan\theta} \frac{F_{max}}{t_1} \int_{t-t_1}^{t} J_1(u) du
\label{eq12}
\end{equation}

The final expression for this phase is given by \hyperref[eq13]{eq.~\ref{eq13}}:

\begin{equation}
h^2(t) = \frac{\pi (1-\nu)}{4 \tan\theta} F_{max} \left[ \frac{1}{G_1} + 
\frac{2t-t_1}{2\eta_1} + \frac{1}{G_2}+ \frac{\tau_2}{G_2 t_1} \left[ e^{- \frac{t}{\tau_2}}-e^{- \frac{t-t_1}{\tau_2}}\right]\right]
\label{eq13}
\end{equation}

The load-displacement curve can be obtained if the viscoelastic properties of materials $J_1(t)$, and $\nu$, are known using the respective 
\hyperref[eq12]{eq.~\ref{eq12}} and \hyperref[eq13]{eq.~\ref{eq13}}. Inversely, if the force-displacement data are obtained experimentally, the 
viscoelastic properties which are the four elements of Burgers model ($G_1$, $G2$, $\eta_1$, $\eta_2$) may be obtained by iterative optimization 
algorithm.

However, one of the easiest ways to test this numerically in ABAQUS (See subsection ~\ref{FEM}) is through Prony series, which is relaxation form of the viscoelastic behavior. Therefore, it is important to determine those coefficients in order to describe the viscoelastic behavior. This is possible through first, determining coefficients of the compliance function $J_1(t)$, then the coefficients of the relaxation function $G(t)$ which is given by \hyperref[eq14]{eq.~\ref{eq14}}

\begin{equation}
G(t) = G_0 \left( 1 - \sum_{i=1}^{n} g_i (1-e^{-\frac{t}{\tau_i}} ) \right)
\label{eq14}
\end{equation}

Where $G_0$ instantaneous shear modulus, $g_i$ relaxation modulus, $\tau_i$ relaxation times.

By determining the compliance function $J_1(t)$, the relaxation function $G(t)$ can be easily determined using Laplace transform of the 
\hyperref[eq3]{eq.~\ref{eq3}} given by \cite{23,24}.

\begin{equation}
\begin{cases}
\displaystyle 2\,\hat{\epsilon}^d(p) = \mathcal{L}(2\,\epsilon^d(t)) = \mathcal{L} \left( \int_0^{t} J_1(t - \tau) 
\frac{d\sigma^d(\tau)}{d\tau} d\tau \right) = p \hat{J_1}(p) \hat{\sigma}^d(p) \\[0.8em]
\displaystyle \hat{\sigma}^d(p) = \mathcal{L}(\sigma^d(t)) = \mathcal{L} \left( \int_0^{t} G(t - \tau) \frac{d\epsilon^d(\tau)}
{d\tau} d\tau \right) = 2 \ p \hat{G}(p) \hat{\epsilon}^d(p)
\end{cases}
\label{eq15}
\end{equation}

Here, \( p \) is the Laplace variable, and \( \hat{\epsilon}^d(p) \), \( \hat{\sigma}^d(p) \), \( \hat{J}_1(p) \), and \( \hat{G}(p) \) are the 
Laplace transforms of \( \epsilon^d(t) \), \( \sigma^d(t) \), \( J_1(t) \), and \( G(t) \), respectively. Therefore:

\begin{equation}
\frac{\hat{\sigma}^d(p)}{2 \,\hat{\epsilon}^d(p)} = p \hat{G}(p) = \frac{1}{p \hat{J_1}(p)}
\label{eq16}
\end{equation}

The relaxation function $G(t)$ can be obtained using inverse Laplace as given by \hyperref[eq17]{eq.~\ref{eq17}}.

\begin{equation}
G(t) = \mathcal{L}^{-1} (\hat{G}(p)) = \mathcal{L}^{-1} \left(\left( {p^{2} \hat{J}_1(p)} \right)^{-1} \right)
\label{eq17}
\end{equation}

Laplace transform of the compliance $J_1(t)$ is given as:

\begin{equation}
\hat{J}_1(p) = \frac{1}{p} \left( \frac{1}{G_1}+\frac{1}{G_2} \right) + \frac{1}{p^2} \frac{1}{\eta_1} - \frac{1}{G_2} \left( \frac{1}{p+\frac{1}{\tau_2}} \right)
\label{eq18}
\end{equation}

Combining \hyperref[eq17]{eq.~\ref{eq17}} and \hyperref[eq18]{eq.~\ref{eq18}}, we find:

\begin{equation}
G(t) = \mathcal{L}^{-1} \left( \left[ \frac{G_1+G_2}{G_1 G_2}p + \frac{1}{\eta_1} - \frac{\tau_2}{G_2} (\frac{p^2}{\tau_2p+1}) 
\right]^{-1} \right)
\label{eq19}
\end{equation}

The inversion of the \hyperref[eq19]{eq.~\ref{eq19}} is performed numerically using Den Iseger algorithm given in the \cite{26} which seems to be more accurate and powerful than the Stehfest algorithm in this case.

\section{FEM-based numerical simulation} \label{FEM}

\begin{figure}[H]
    \centering
    \includegraphics[width=0.8\linewidth]{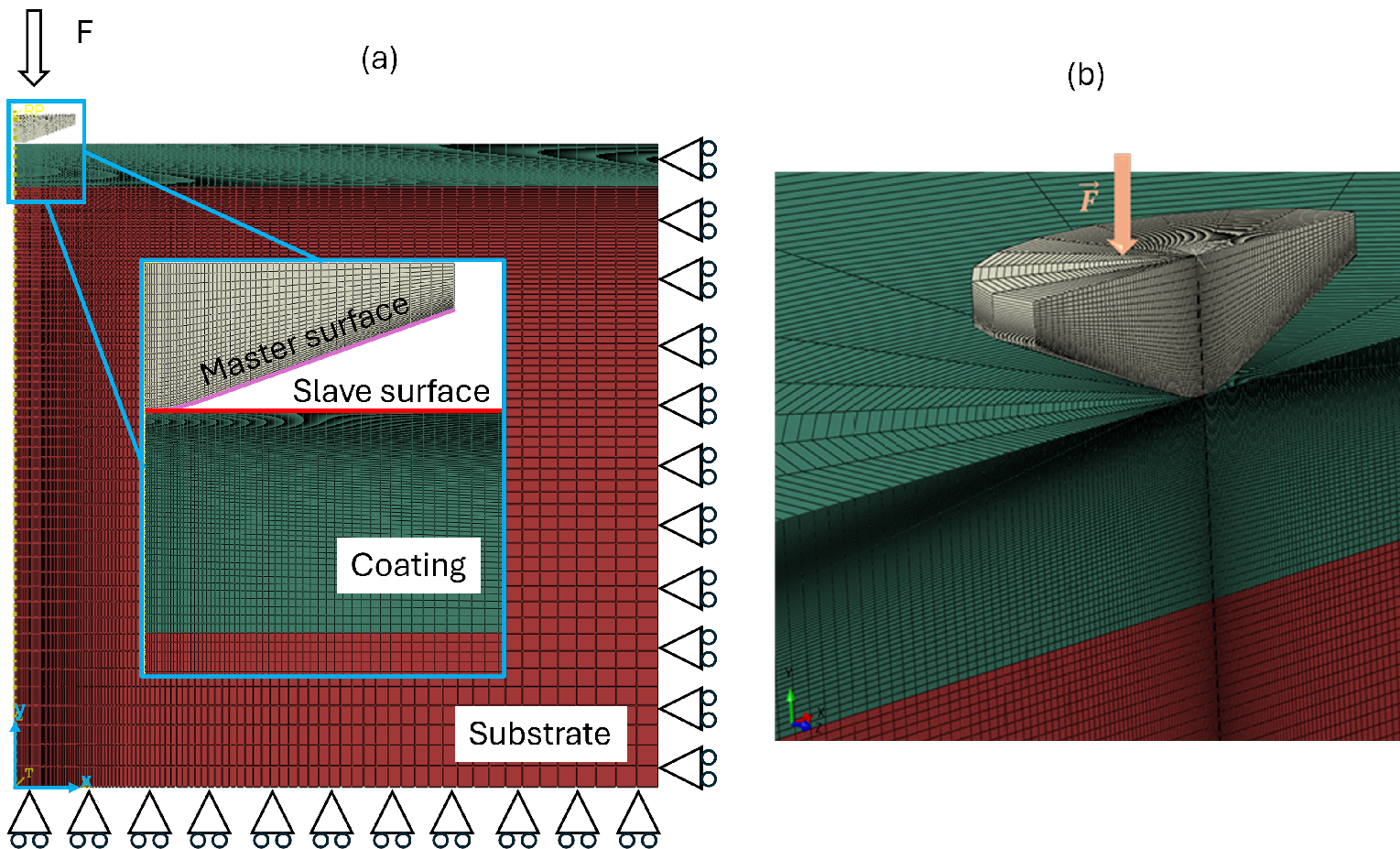}
    \caption{A 2D axisymmetric FEM model used in the simulations (a) illustrating coating/substrate with applied boundary conditions, (b) 3D geometric visualization obtained by revolving the axisymmetric section around the symmetry axis showing a refined mesh with element sizes ranging from 500 nm to 10 nm. The mesh consists of 21,001 CAX4R elements and 21,368 nodes, of which 19,516 belong to the sample. The computational time for a single simulation is approximately 150 s, while the complete optimization required about 24 hours on four processing cores.}
    \label{fig3}
\end{figure}

As illustrated in \hyperref[fig3]{Fig.~\ref{fig3}}, the model built in the FEM software ABAQUS is a simplified two-dimensional axisymmetric model with a radius and height of 15 \textmu m. The finite element model was developed using a two-dimensional axisymmetric formulation to reduce computational cost. This approach is justified by approximating the three-dimensional Berkovich indenter with an equivalent conical indenter having a semi-apex angle of $70.3^\circ$, thereby preserving the projected contact area of the original Berkovich geometry \cite{16,27}. The contact is assumed to be perfect between the coating of 1 \textmu m depth and a steel substrate of 14 \textmu m depth. The indenter is represented as a rigid conical body with the mechanical properties of diamond ($E_{\text{indenter}}$ = 1141 GPa, $\nu = 0.07$). Preliminary studies were conducted to examine the influence of sample dimensions and tip radius showed that the sample dimensions should be at least 7 \textmu m, and the tip radius was determined to be 660 nm, which is in the same order of magnitude as the 809 nm found by \cite{28}. The steel substrate is assumed to behave as a perfect elastoplastic material (no strain hardening is introduced) where the coating is assumed to be viscoelastic material. The mechanical properties are summarized in table \ref{tab2}.

\begin{table}[htbp]
\centering
\captionsetup{justification=raggedright, singlelinecheck=false}
\caption{Material properties of different parts.}
\label{tab2}
\resizebox{\textwidth}{!}{%
\begin{tabular}{lccccc}
\toprule
Material Name & Young's modulus (GPa) & Poisson's ratio ($\nu$) & Density (kg/m$^3$) & Yield stress (GPa) & Thickness (\textmu m) \\
\midrule
Diamond              & 1141      & 0.07 & 3510 & --  & -- \\
WSe\textsubscript{2} & TBE$^{*}$ & 0.19~\cite{v5} & 9320 & --  & 1 \\
Steel substrate      & 200       & 0.28 & 7850 & 1.9 & 14 \\
\bottomrule
\end{tabular}%
}
{\raggedright \footnotesize $^{*}$TBE: To be estimated.\par}
\end{table}

As illustrated in \hyperref[fig3]{Fig.~\ref{fig3}}, the sample is fully constrained along the lateral and the bottom surfaces in the normal directions to those surface as well as in rotation. The load $F$ (or pressure $P$ in ABAQUS) is applied on the top surface of the indenter toward the sample along negative direction of $Z$. This load corresponds to the experimental history $H(t)$ with a maximum load $F_{max}$ (\hyperref[eq1]{eq.~\ref{eq1}}) allows the obtaining of displacement response at the hold phase. A tangential behavior with a penalty coefficient of 0.07, according to previous study on tribological properties of WSe\textsubscript{2} coating \cite{14}, is applied to the 
contact properties between the indenter and the sample’s upper surface. The interaction is defined as a surface-to-surface contact in ABAQUS standard, with the indenter’s lower surface as the master surface and the sample’s upper surface as the slave surface. Finally, the dynamic implicit step is employed due to the time-dependent nature of the problem.

In the meshing process, the used mesh type is the axisymmetric bilinear quadrilateral elements with reduced integration (CAX4R). The elements are seeded with a size-based bias, starting at 500 nm for the elements farthest from the contact point situated at the lateral surface and reducing to 10 nm near the first contact point. For the longitudinal direction, the meshing is refined to 1 nm at the contact and coarsened to 42 nm at the interface coating/substrate, then from 42 nm to 500 nm at the bottom surface. As a result, the element sizes decrease as they approach the contact region at the sample's upper surface and the axisymmetric axis which allows to capture the material behavior around the 
zone of contact effectively without unnecessary computational cost elsewhere which is expensive.

\section{Optimization process} \label{Optimization process}

\subsection{Nelder-Mead simplex algorithm} \label{Nelder-Mead simplex algorithm}

Optimization algorithms are commonly used tools in various engineering and scientific applications, including calibrating finite element models 
and estimating material properties. These algorithms aim to minimize or maximize a given objective function by iteratively adjusting the 
variables. Among the most widely used optimization algorithms are gradient-based methods, evolutionary algorithms, and direct search methods 
like the Nelder-Mead simplex algorithm \cite{29}.

First-order methods, also known as gradient-based methods like steepest descent and conjugate gradient, rely on calculating the gradients (or 
derivatives) of the objective function with respect to the parameters. These algorithms are typically fast and efficient when the objective 
function is smooth and differentiable. However, the main drawback is their tendency to get trapped in local minima, making them less effective 
when dealing with noisy, discontinuous, or non-differentiable functions. In contrast, second-order methods, such as Newton's method and quasi-
Newton methods, provide faster convergence by incorporating curvature information from the second derivative (Hessian matrix). Newton's method 
uses both the Jacobian and the Hessian matrices to determine the direction and size of each step, resulting in rapid convergence, especially 
near the solution. However, calculating the Hessian can be computationally expensive for high-dimensional problems, and the method can become 
unstable if the Hessian is poorly conditioned.

\begin{figure}[H]
    \centering
    \includegraphics[width=0.8\linewidth]{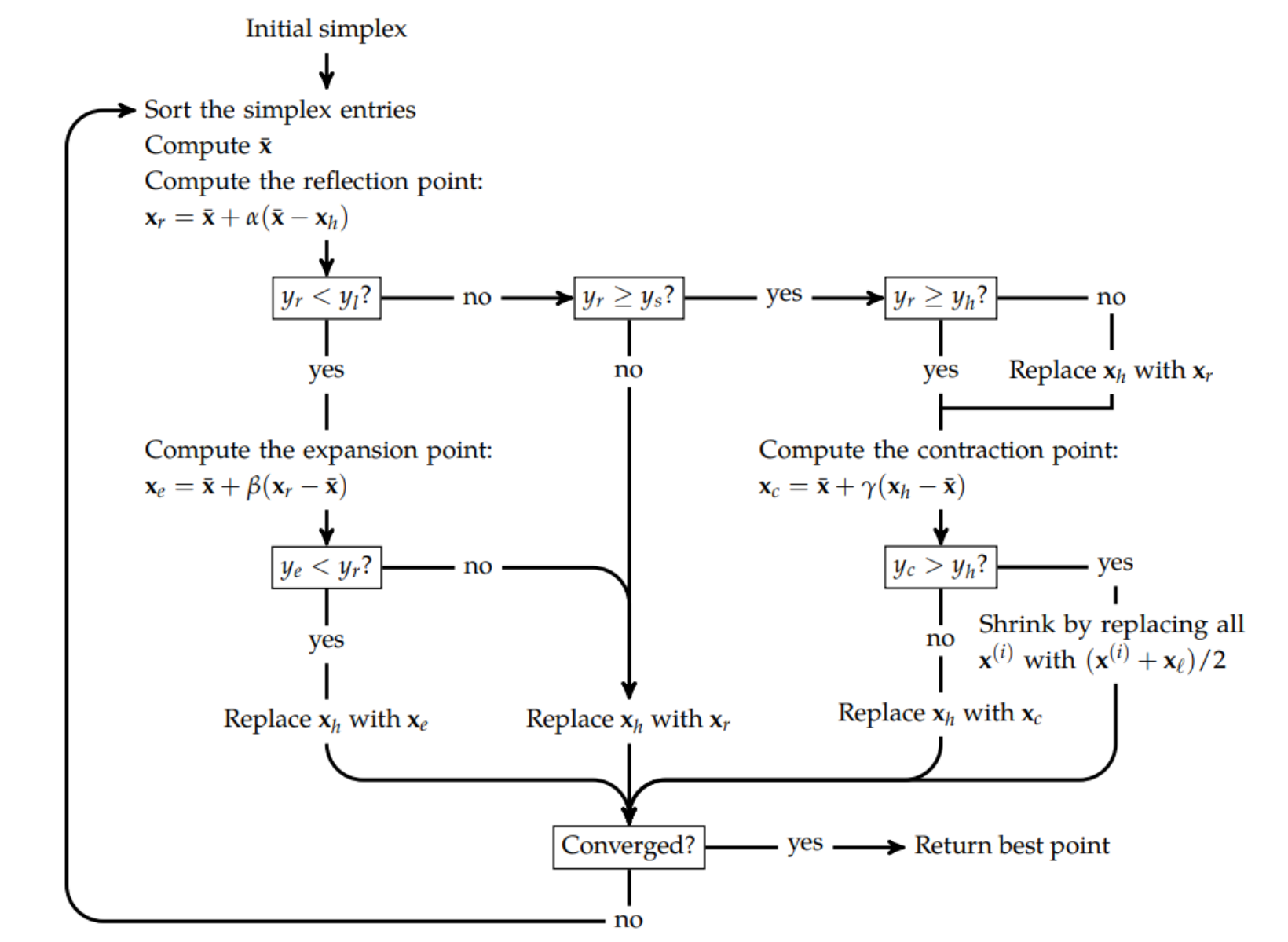}
    \caption{Nelder-Mead method flowchart illustrated in \cite{30}.}
    \label{fig4}
\end{figure}

In this study, a different approach, called a direct search method (also referred to as zero-order, black-box or derivative-free), is employed 
to estimate the mechanical properties of irradiated samples. Specifically, the Nelder-Mead simplex method is used \cite{29,30}. Unlike gradient-
based methods, the Nelder-Mead method does not require any gradient or Hessian information to guide the optimization process. Instead, it uses 
a simplex, a geometric shape formed by evaluating the objective function at its vertices, to search for the minimum \cite{30}. The method 
incrementally adjusts the Simplex's shape and step size as the search progresses. Its main advantage lies in its simplicity and versatility, 
making it suitable for a wide range of problems, particularly those with non-differentiable objective functions or those derived from 
simulations where gradients are not easily accessible. The Nelder-Mead method follows a set of rules that guide how the simplex is updated, 
based on the evaluations of the objective function at its vertices \cite{30}. The process is illustrated in a flowchart in \hyperref[fig4]
{Fig.~\ref{fig4}}.

Additional details on the operations of the Nelder–Mead algorithm, illustrating how the simplex evolves and adapts during the search for the optimal solution, are provided in Appendix \ref{Appendix A}.

\subsection{Scheme of the computation} \label{Scheme of the computation}

In this study, the Nelder–Mead optimization algorithm was employed to identify the viscoelastic material parameters, as described in Section~\ref{Nelder-Mead simplex algorithm}. This approach differs from the approach outlined in \cite{31,32}, which relies on a linear function relating mechanical properties to the maximum load $F$ and stiffness $S$. It was developed after performing direct trial FEM simulations with the values that bracket the expected material properties. This method may focus on maximum load alone and limit the consideration of full experimental data and its potential impact on the accuracy of the material property estimation.

In contrast, our approach proposes using the entire dataset of displacement and force, rather than relying solely on maximum load and stiffness. Following this approach requires the development of several functions (in this study, PYTHON was used) to exchange data with and full control of ABAQUS to establish the optimization process.

\begin{figure}[H]
    \centering
    \includegraphics[width=0.8\linewidth]{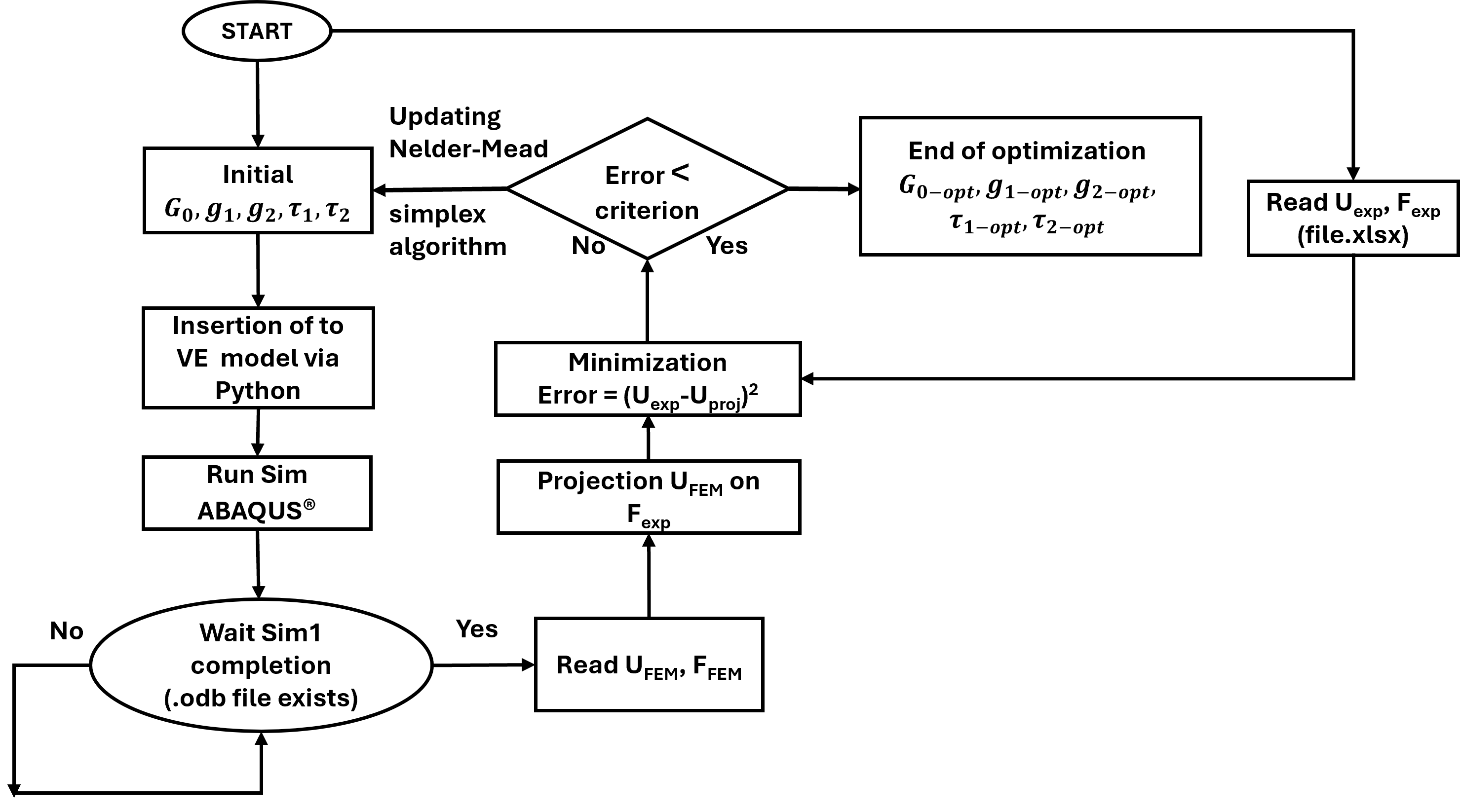}
    \caption{Automated optimization algorithm of elastic-viscoelastic parameters implemented in PYTHON.}
    \label{fig5}
\end{figure}

The optimization workflow shown in \hyperref[fig4]{Fig.~\ref{fig5}} is employed to simultaneously determine the viscoelastic properties of the coating. This is through fitting the output data of FE simulation to the experimental data. The entire process from reading the experimental data, updating the properties using the simplex algorithm, running and monitoring the generation of output files (e.g., the .odb file) within ABAQUS, is fully automated and managed through a PYTHON script. Initially, the script inserts an initial guess for properties. Upon completion of the first simulation (via continuous checking of the existing ODB file), the script extracts the simulated force-displacement data. To 
ensure consistent comparison, the simulated force is projected onto the experimental displacement data, and the sum of squared errors between 
the experimental and projected force values is evaluated against a predefined convergence criterion.

If the convergence criterion is not met, the parameters to be estimated (in our case the Prony series $G_0$, $g_i$, and $\tau_i$) are 
updated using the optimization algorithm, and the loop is repeated. This iterative process continues until the error falls below the defined 
threshold. Upon convergence, optimization yields the final calibrated values: $G_{0-optim}$, $g_{i-optim}$ and $\tau_{i-optim}$. The evolution of the estimated viscoelastic properties and the corresponding optimization error are presented in \hyperref[fig9]{Fig.~\ref{fig9}}.

\section{Results and discussion} \label{results}

\subsection{Tested Material} \label{Material}

Thin film of WSe\textsubscript{2} was fabricated at temperature of 300 °C using an AJA magnetron sputtering system operated with a 150 W DC power supply and pure WSe\textsubscript{2} targets. The deposition was carried out at a base vacuum of $5\times10^{-3}$ Pa and a working pressure of 0.67 Pa for 7200 s, resulting in a coating thickness of approximately 1030 nm, Polished tool steel substrates (WNr. 1.2379, hardness \(\approx\) 9 GPa) were employed for the depositions. Prior to coating, the substrates were cleaned by RF sputtering at 50 W for 30 min. To promote adhesion, a chromium interlayer approximately 150 nm thick was first deposited between the substrate and the coating. Coating thickness was evaluated using a Zygo NewView 8000 profilometer.

\subsection{Analytical modeling results} \label{Analytical results}

\hyperref[fig7]{Fig.~\ref{fig7}} presents the experimental force-displacement curves obtained from 9 representative nanoindentation tests on 
WSe\textsubscript{2}, and the corresponding averaged curve. The averaged indentation curve shows a maximum penetration depth of 
approximately 98.2 nm at peak load. During the subsequent 2 s hold period, the displacement increases from 95.9 nm to 98.2 nm, revealing local 
viscoelastic behavior characterized by an indentation creep of 0.024.

\begin{figure}[H]
    \centering

    \begin{subfigure}[c]{0.45\linewidth}
        \centering
        \includegraphics[width=\linewidth]{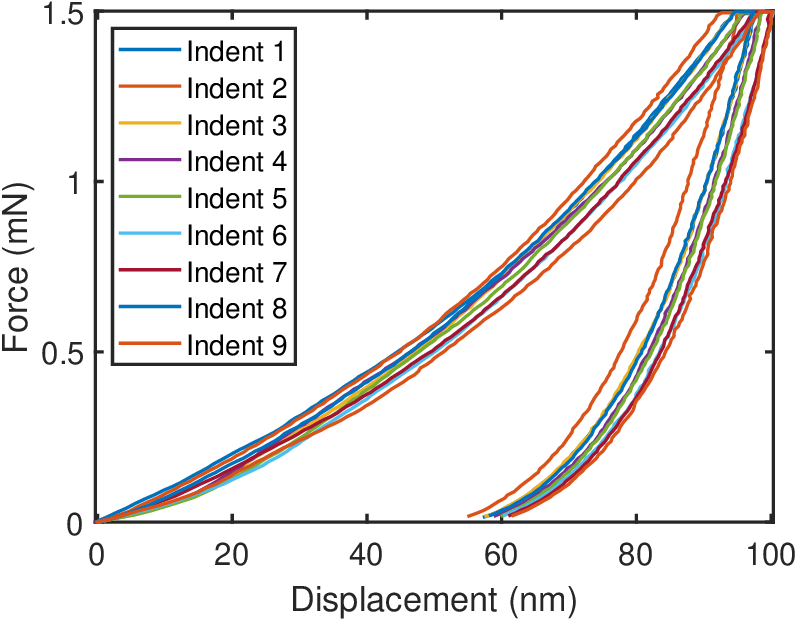}
        \caption{}
        \label{fig6a}
    \end{subfigure}
    \hfill
    \begin{subfigure}[c]{0.45\linewidth}
        \centering
        \includegraphics[width=\linewidth]{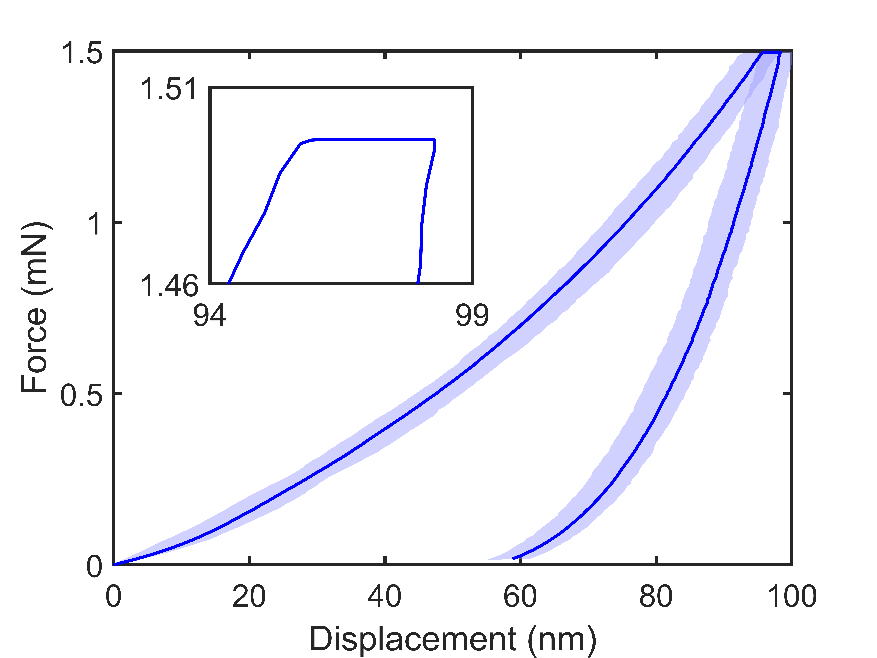}
        \caption{}
        \label{fig6b}
    \end{subfigure}

    \caption{(a) Experimental force-displacement curves, (b) Averaged force-displacement curve including the hold phase with the error band.}
    \label{fig6}
\end{figure}

As previously discussed, the viscoelastic model is applicable only during the loading and holding phases, as well as in the initial portion of the unloading phase where the contact area continues to increase monotonically \cite{24,22,23}. Consequently, the constitutive formulation described in Section~\ref{Constitutive model} was applied exclusively to the loading and hold phases. The displacement expressions defined in 
\hyperref[eq10]{Eq.~\ref{eq10}} and \hyperref[eq13]{Eq.~\ref{eq13}} were combined with the experimental loading function to develop a nonlinear 
least-squares curve-fitting implemented in MATLAB. This approach minimizes the quadratic error between the experimental and the simulated displacement during the load and the hold phases, which enables the estimation of the viscoelastic (creep) parameters $G_1$, $G_2$, $\eta_1$, and $\eta_2$. The fitting results are presented in \hyperref[fig7]{Fig.~\ref{fig7a}}. The optimization process was initialized with the following parameter values: $G_1 = 1$~GPa, $G_2 = 1$~GPa, $\eta_1 = 1$~GPa·s, and $\eta_2 = 1$~GPa·s.

In addition to the analytical Burgers model, the elastic properties of the coating were independently evaluated using the classical Oliver–Pharr (OP) method. This approach determines the elastic response from the initial unloading stiffness of the nanoindentation curve while accounting for the geometrical correction factor associated with the Berkovich indenter. The unloading segment was first fitted using a power-law relationship to determine the initial unloading stiffness. The contact depth, projected contact area, hardness, reduced elastic modulus, Young's modulus, and finally the shear modulus were then successively calculated according to:

\begin{equation}
\left\{
\begin{aligned}
S &= \left.\frac{dP}{dh}\right|_{h=h_{\max}},
\,\,\,\,\,\,\,\,\,\,\,\,\,h_c = h_{\max}-\varepsilon\frac{P_{\max}}{S}\\[0.5em]
A_c &= f(h_c),
\,\,\, \,\,\,H = \frac{P_{\max}}{A_c},
\,\,\,\,\,\,\,\,E_r = \frac{\sqrt{\pi}}{2\beta}\frac{S}{\sqrt{A_c}}\\[0.5em]
E &= \left(\frac{1}{E_r}-\frac{1-\nu_i^2}{E_i}\right)^{-1}(1-\nu^2),
\,\,\,\,\,\,\,\,\,\,\,\,G = \frac{E}{2(1+\nu)}
\end{aligned}
\right.
\label{eq:OP}
\end{equation}

Where $S$ is the initial unloading stiffness, $h_c$ is the contact depth, $\varepsilon$ is the geometrical constant for a Berkovich indenter, $A_c$ is the projected contact area determined from the calibrated area function, $H$ is the hardness, $E_r$ is the reduced elastic modulus, $\beta=1.05$ is a correction factor, $E_i$ and $\nu_i$, are the Young's modulus and Poisson's ratio of the diamond indenter, $E$ and $\nu$ are the Young's modulus and Poisson's ratio of the tested coating, respectively, and $G$ is the corresponding shear modulus of the coating.

It is worth noting that the shear modulus obtained from the classical OP method corresponds to an apparent unloading modulus extracted from the initial unloading stiffness. However, because the material has already undergone viscoelastic deformation during the loading and hold phases, the measured unloading stiffness is affected by the time-dependent deformation, leading to an underestimation of the elastic properties. To account for this effect, Ngan and co-workers proposed a correction that enables the extraction of the intrinsic elastic properties from the viscoelastic nanoindentation response \cite{108,112,113,114}. This correction incorporates both the displacement rate at the end of the hold period and the unloading rate to determine the corrected unloading stiffness according to:

\begin{equation}
\frac{1}{S_{e}} = \frac{1}{S} - \frac{\dot{h}_{h}}{\dot{P}_{u}}
\label{eq21}
\end{equation}

where $\dot{h}_{h}$ denotes the displacement rate at the end of the hold period immediately before unloading, and $\dot{P}_{u}$ is the initial unloading rate. Using the corrected unloading stiffness, the reduced elastic modulus can then be expressed as:

\begin{equation}
E_{r} = E_r = \frac{\sqrt{\pi}}{2\beta}\frac{1}{\sqrt{A_c}\,\left(\frac{1}{S} - \frac{\dot{h}_{h}}{\dot{P}_{u}} \right)}
\label{eq22}
\end{equation}

The displacement rate at the end of the hold period is obtained by differentiating the Burgers displacement equation (\hyperref[eq13]{Eq.~\ref{eq13}}) with respect to time, giving:

\begin{equation}
\dot{h}_{h} = \left.\dot{h} \right|_{t=t_{\text{final-hold}}} = \frac{1}{2h(t_{\text{final-hold}})}\frac{dh^{2}(t_{\text{final-hold}})}{dt}
\label{eq23}
\end{equation}

As indicated by \hyperref[eq21]{Eq.~\ref{eq21}} and \hyperref[eq22]{Eq.~\ref{eq22}}, the classical OP method underestimates the unloading stiffness and, consequently, the elastic modulus of viscoelastic materials. By applying the proposed correction, the contribution of the time-dependent viscoelastic deformation is taken into account, leading to a more accurate determination of the elastic properties. The corrected unloading fit and the corresponding mechanical properties are presented in \hyperref[fig7]{Fig.~\ref{fig7b}}.

\begin{figure}[H]
    \centering

    \begin{subfigure}[c]{0.45\linewidth}
        \centering
        \includegraphics[width=\linewidth]{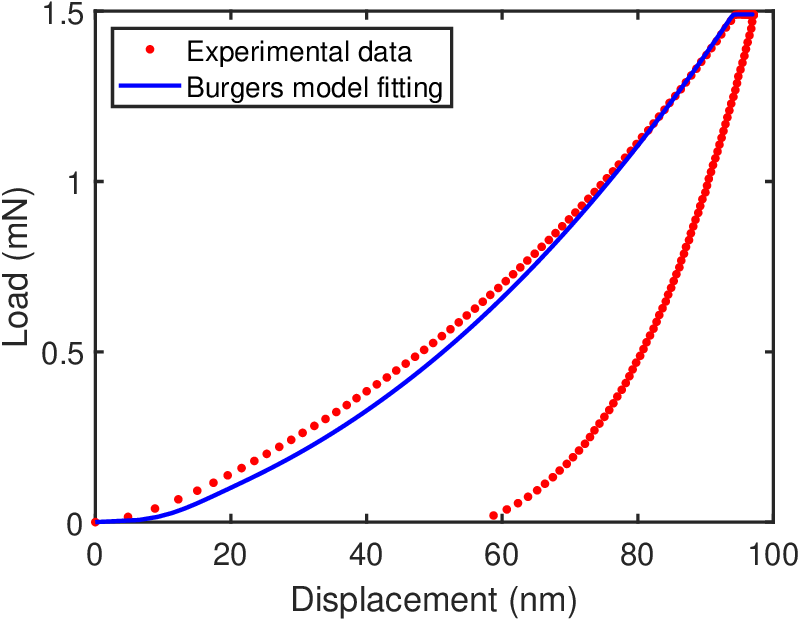}
        \caption{}
        \label{fig7a}
    \end{subfigure}
    \hfill
    \begin{subfigure}[c]{0.45\linewidth}
        \centering
        \includegraphics[width=\linewidth]{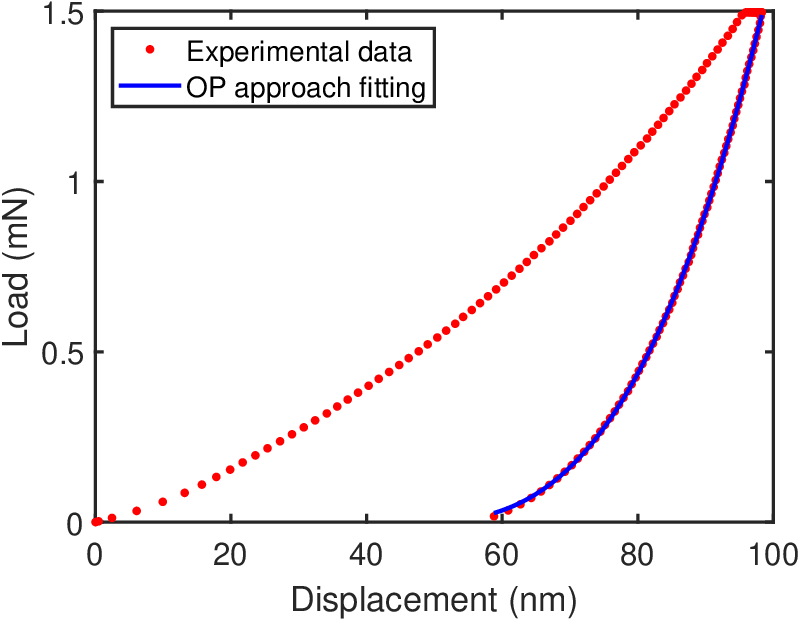}
        \caption{}
        \label{fig7b}
    \end{subfigure}

    \caption{Fit of experimental force-displacement data (a) during the load and hold phases using Burgers chain for Viscoelastic model using optimal parameters shown in table 3, (b) during the unload phase using Oliver-Pharr approach using optimal parameters listed in table 3.}
    \label{fig7}
\end{figure}

During the loading and hold phases of nanoindentation, the Burgers viscoelastic model effectively captures the material response, particularly during the hold phase and the later stage of loading, where the deformation is gradual and time-dependent creep dominates. In these regions, the model accurately reproduces both the elastic recovery and the viscous flow of the coating. The viscoelastic parameters are identified by fitting the hold segment of the nanoindentation curve, where the applied load remains constant. Nevertheless, as shown by \hyperref[eq13]{Eq.~\ref{eq13}}, the displacement during the hold phase depends on the complete loading history through the convolution integral. Consequently, the parameter identification is not based solely on the hold response but also incorporates the effect of the preceding loading stage. Furthermore, according to \cite{22}, the model remains valid during the initial stage of unloading, as long as the contact area continues to increase monotonically. This condition is satisfied throughout the loading, hold, and early unloading stages of the present nanoindentation test, allowing the model to describe the initial unloading response as well. However, during the initial stage of loading, the material response is essentially instantaneous and more complex than can be represented by the Burgers model. Since the Burgers model consists of only two springs and two dashpots, it cannot account for additional phenomena such as inertial and surface effects, which may contribute to the slight mismatch observed during the early loading stage (see, for example, Fig. 4.9 in \cite{23}). Despite this limitation, the model provides an excellent overall agreement with the experimental data, accurately reproducing the coating response throughout the loading and hold phases.

At the early stage of loading, the short-time behavior of the Burgers creep compliance in \hyperref[eq6]{eq.~\ref{eq6}} is $\left[ \frac{1}{G_1}+t\left(\frac{1}{\eta_1}+\frac{1}{\eta_2}\right) \right]$
showing an instantaneous elastic contribution followed by a linear viscous contribution. Therefore, introducing an Abel dashpot, as in \cite{v3} $(\sigma_{\mathrm{dashpot}}=\eta^\alpha d^\alpha\varepsilon/dt^\alpha)$ instead of ($\sigma_{\mathrm{dashpot}} = \eta d\epsilon/dt$), may improve the description of the fast elastic-viscoelastic transition at the early loading stage from $\left[ \frac{1}{G_1} + t \left ( \frac{1}{\eta_1}+\frac{1}{\eta_2} \right) \right]$ to $\left[ \frac{1}{G_1}+ \left(\frac{1}{\eta_1^\alpha}+\frac{1}{\eta_2^\alpha}\right) \frac{t^\alpha}{\Gamma(1+\alpha)}\right]$.

The optimal viscoelastic parameters identified using the Burgers model, together with the elastic properties obtained from the OP method, are summarized in Table~\ref{tab-params}.

\begin{table}[H]
\captionsetup{justification=raggedright, singlelinecheck=false}
\caption{Optimized parameters of the creep function $J(t)$ and the Oliver–Pharr (OP) model.}
\label{tab-params}
\resizebox{\textwidth}{!}{%
\begin{tabular}{lcccccccc}
\toprule
\multicolumn{8}{c}{\textbf{Creep function $J(t)$ parameters}} \\
\midrule
Parameter & $G_1$ (GPa) & $\eta_1$ (GPa·s) & $G_2$ (GPa) & $\eta_2$ (GPa·s) & RMSE & $R^2$ & Iterations \\
\midrule
Value & 55.5 & 6702 & 110 & 89.2 & 0.0026 & 0.997 & 38 \\
\midrule
\multicolumn{8}{c}{\textbf{Oliver–Pharr (OP) model parameters}} \\
\midrule
Parameter & $S$ (N·\textmu m$^{-1}$) & $S_{e}$ (N·\textmu m$^{-1}$) & $h_c$ (\textmu m) & $E_r$ (GPa) & $G$ (GPa) & $H$ (GPa) & RMSE & $R^2$ \\
\midrule
Value & 0.081 & 0.104 & 0.087 & 113.61 & 51.10 & 2.49 & 0.0031 & 0.999 \\
\bottomrule
\end{tabular}%
}

\end{table}

The remaining discrepancy can mainly be attributed to the different fitting procedures. As shown in \hyperref[fig7a]{Fig.~\ref{fig7a}}, the Burgers model exhibits a slight mismatch during the initial loading stage, where the experimental indentation response does not evolve linearly with time as predicted by the classical Burgers model. This behavior suggests that the short-time viscoelastic response is more complex than can be represented by the two-dashpot Burgers formulation and could potentially be better described by a fractional viscoelastic model (Abel dashpot), in which the creep evolves as $t^\alpha (0<\alpha<1)$. In contrast, the OP method determines the unloading stiffness from a power-law fit of the unloading segment, providing an excellent agreement with the experimental data as shown in \hyperref[fig7b]{Fig.~\ref{fig7b}}. Nevertheless, the corrected OP modulus also depends on the displacement rate at the end of the hold period, $\dot{h}_{h}$, which is evaluated using the Burgers model. Consequently, the corrected modulus is indirectly influenced by the Burgers approximation. Despite these differences, both approaches provide consistent and reliable estimation of the elastic properties of the viscoelastic coating.

\begin{figure}[H]
    \centering
    \includegraphics[width=0.6\linewidth]{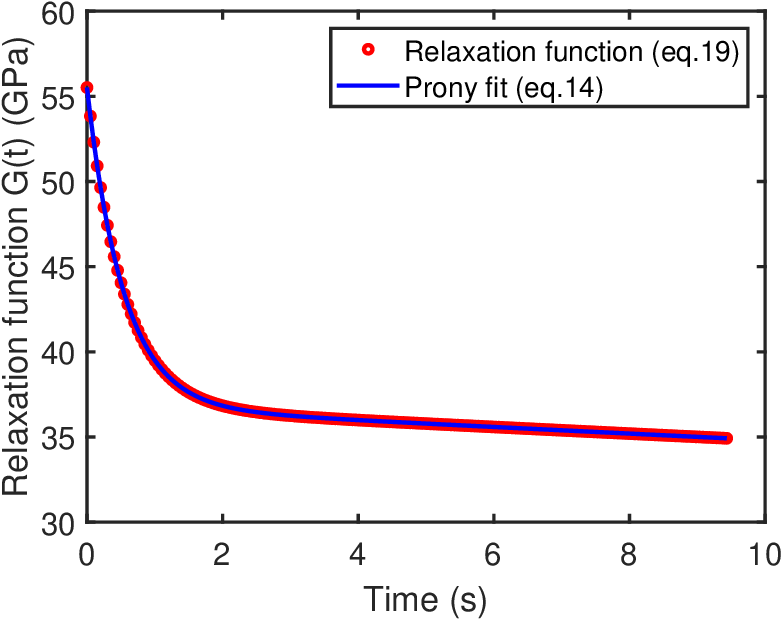}
    \caption{Comparison between the relaxation function $G(t)$ obtained from the inverse Laplace transform \hyperref[eq19]{Eq.~\ref{eq19}} and its equivalent two-term Prony-series approximation \hyperref[eq14]{Eq.~\ref{eq14}}.}
    \label{fig8}
\end{figure}

Furthermore, to compare the analytical Burgers model with the FEM-based Prony formulation, the optimized Burgers creep parameters were converted into an equivalent relaxation form. The relaxation function $G(t)$ was first obtained from the creep compliance using the inverse Laplace transform given in \hyperref[eq19]{Eq.~\ref{eq19}}. Then, the resulting relaxation function was approximated by the two-term Prony-series expression given in \hyperref[eq14]{Eq.~\ref{eq14}} through nonlinear least-squares fitting in MATLAB. The comparison between the relaxation function obtained from \hyperref[eq19]{Eq.~\ref{eq19}} and its Prony-series approximation is shown in \hyperref[fig8]{Fig.~\ref{fig8}}.

The optimal Prony parameters obtained from the analytical-model fitting were $G_0 = 55.5$ GPa, $E = 132$ GPa, $g_1 = 0.34$, $g_2 = 0.62$, $\tau_1 = 0.54$ s, and $\tau_2 = 170.4$ s. The Prony-series approximation accurately reproduced the relaxation function, yielding an RMSE of 0.0001 and an $R^2$ value of 1.000 after 170 iterations. In the following section, these analytically derived Prony parameters are compared with those independently identified through the finite element inverse optimization based on the Nelder–Mead simplex algorithm.

\subsection{Numerical modeling results} \label{Numerical results}

\begin{figure}[H]
    \centering
    \begin{subfigure}[c]{0.3\linewidth}
        \centering
        \includegraphics[width=\linewidth]{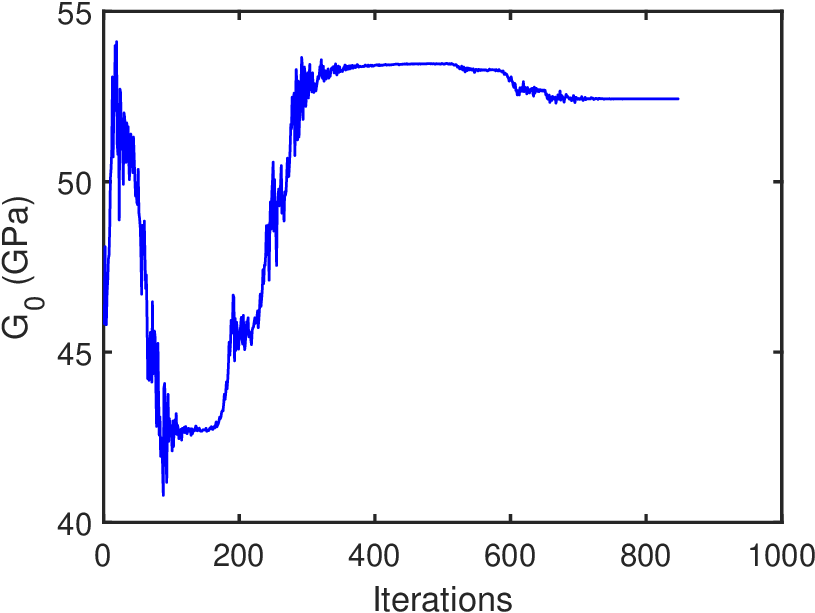}
        \caption{}
        \label{fig9a}
    \end{subfigure}
    \hfill
    \begin{subfigure}[c]{0.3\linewidth}
        \centering
        \includegraphics[width=\linewidth]{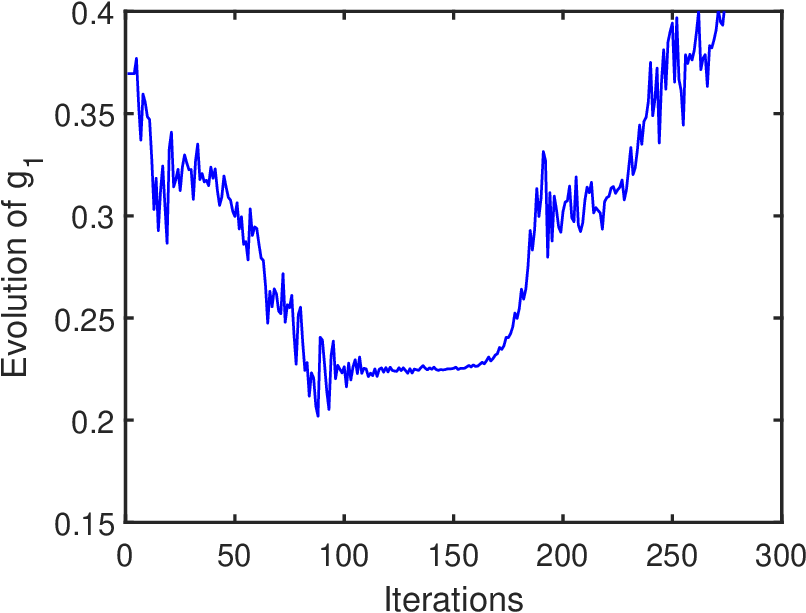}
        \caption{}
        \label{fig9b}
    \end{subfigure}
    \hfill
    \begin{subfigure}[c]{0.3\linewidth}
        \centering
        \includegraphics[width=\linewidth]{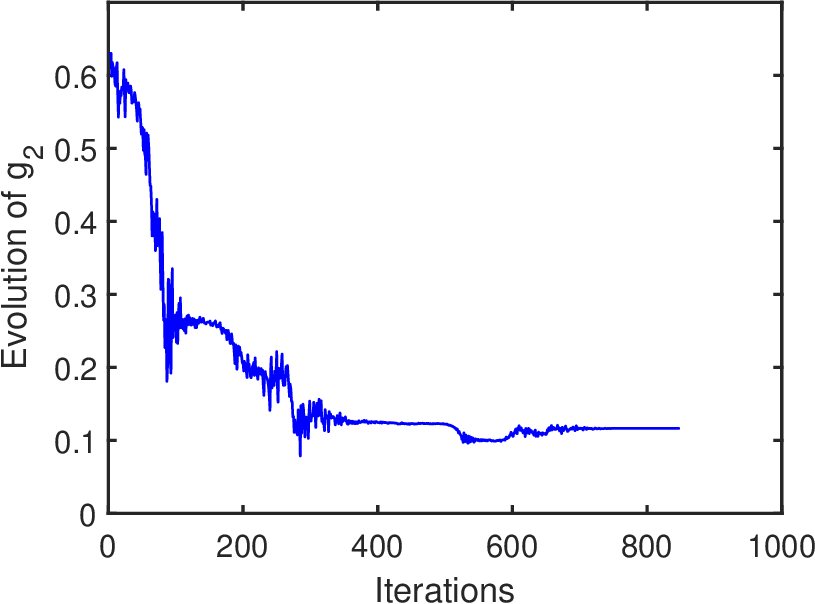}
        \caption{}
        \label{fig9c}
    \end{subfigure}

    \vspace{0.5em} 

    \begin{subfigure}[c]{0.3\linewidth}
        \centering
        \includegraphics[width=\linewidth]{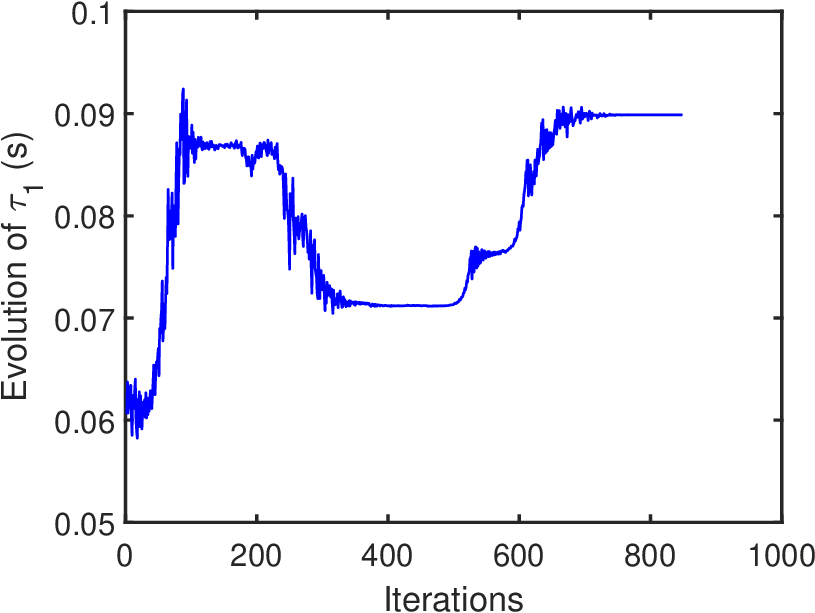}
        \caption{}
        \label{fig9d}
    \end{subfigure}
    \hfill
    \begin{subfigure}[c]{0.3\linewidth}
        \centering
        \includegraphics[width=\linewidth]{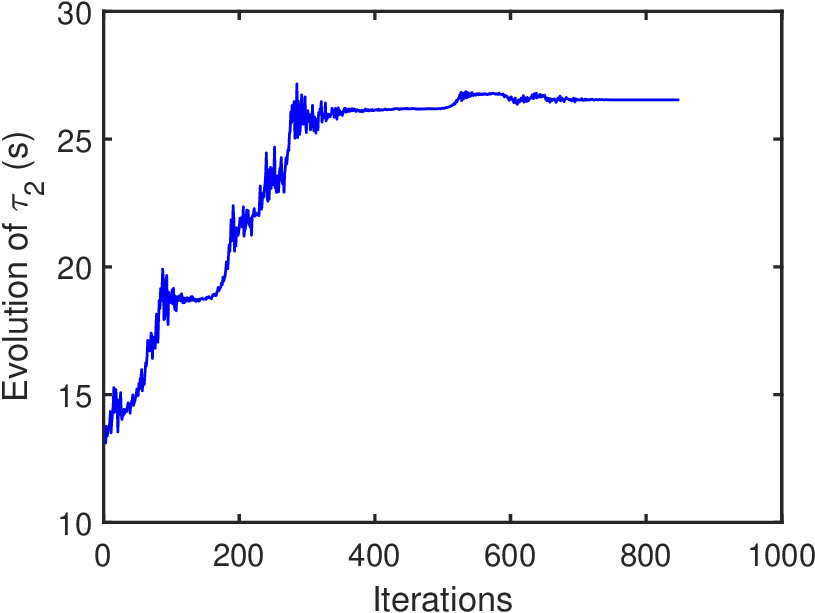}
        \caption{}
        \label{fig9e}
    \end{subfigure}
    \hfill
    \begin{subfigure}[c]{0.3\linewidth}
        \centering
        \includegraphics[width=\linewidth]{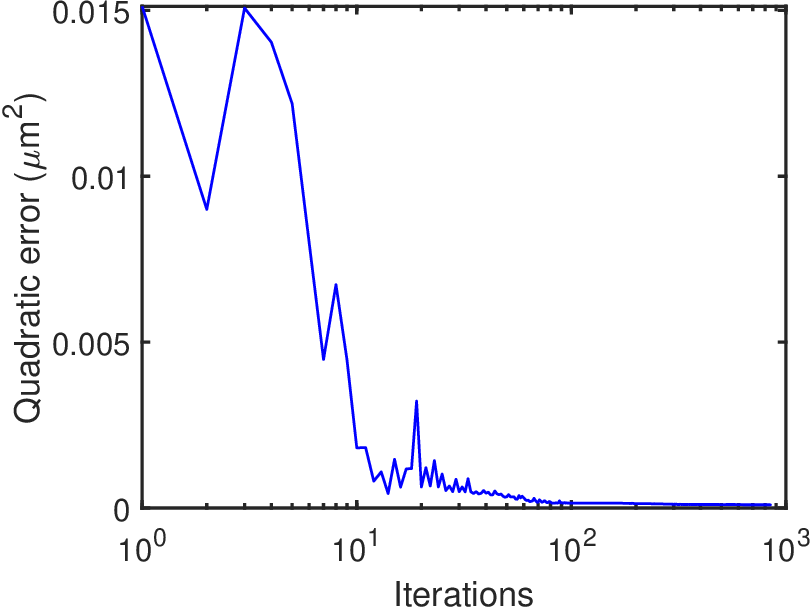}
        \caption{}
        \label{fig9f}
    \end{subfigure}

    \caption{Evolution of the elastic and viscoelastic parameters over the iterations during the optimization process minimizing the error 
    between the experimental and FEM-fitted displacements: (a) shear modulus $\left(G_0 = \frac{E}{2(1+\nu)}\right)$, (b) first relaxation 
    constant, (c) second relaxation constant, (d) first relaxation time, (e) second relaxation time, and (f) quadratic error (sum of squared 
    errors, SSE).}

    \label{fig9}
\end{figure}

As described in Section~\ref{FEM}, a 2D axisymmetric viscoelastic finite element model was developed to estimate the Young's modulus and viscoelastic parameters (Prony series coefficients), in a manner consistent with the analytical approach. The model was reduced to two dimensions while preserving the contact area by using the appropriate equivalent cone angle of the indenter. This simplification was made to accelerate the simulations while maintaining accuracy and comparability with experimental results, as the estimation of viscoelastic parameters requires iterative simulations, which would be computationally expensive in a full 3D model. Therefore, similarly to the previous section 
(\ref{Analytical results}), the relaxation coefficients are estimated by minimizing the quadratic error between the experimental and simulated 
displacement data. The optimization is performed in PYTHON using the minimize function from the SciPy library, based on the Nelder–Mead method. This method is well-suited for this problem as it employs a simplex-based search for local minima and does not require gradient information, which is advantageous given that the finite element model provides only an output and not an explicit, differentiable formulation. The initial guess of parameters are taken randomly but close to the optimal parameters found in analytical model-based optimization. The obtained results of optimization using numerical model within ABAQUS are illustrated 
in \hyperref[fig9]{Fig.~\ref{fig9}}.

We briefly outline the evolution of the estimated parameters during the optimization process. The Young's modulus, initially set to 108~GPa, 
progressively stabilizes around 123.7~GPa. Similarly, the relaxation parameters $g_1$ and $g_2$, which were initially assigned values of 0.37 and 0.63, converge to final values of 0.44 and 0.12, respectively. The relaxation times $\tau_1$ and $\tau_2$, starting from 0.06~s and 13~s, converge to 0.09~s and 26.5~s. These trends demonstrate the efficiency of the optimization procedure in refining both the elastic and viscoelastic parameters to closely match the experimental displacement, as reflected by the reduction in error shown in \hyperref[fig9]{Fig.~\ref{fig9f}}.

The optimal Prony-series parameters identified through the finite element inverse optimization were $G_0 = 52$ GPa, $E = 123.7$ GPa, $g_1 = 0.44$, $g_2 = 0.12$, $\tau_1 = 0.09$ s, and $\tau_2 = 26.5$ s. The optimization converged after 848 iterations, yielding an RMSE of $7.12 \times 10^{-4}$ and $R^2$ of 0.99.

Although both the analytical and numerical approaches reproduce the experimental nanoindentation response with excellent accuracy, the identified instantaneous shear modulus is in good agreement with the value obtained from the corrected Oliver-Pharr method, confirming the consistency of the three approaches in estimating the elastic properties of the coating. In contrast, the remaining viscoelastic parameters, particularly the characteristic relaxation times, differ between the analytical Burgers model and the finite element Prony-series model. This discrepancy mainly arises because the two approaches rely on fundamentally different mechanical formulations. The analytical model represents the material by an equivalent rheological network subjected to homogeneous loading conditions and therefore provides a lumped macroscopic description of the viscoelastic behavior. In contrast, the finite element model solves the continuum contact problem, accounting for the evolution of the contact area together with the heterogeneous stress and strain fields that develop beneath the indenter. Consequently, the Prony-series parameters identified by the FEM represent an equivalent description of the localized indentation response, whereas the Burgers parameters characterize an equivalent macroscopic viscoelastic behavior.

 The axisymmetric finite element formulation may also contribute to the observed differences. The 2D model assumes circumferential symmetry of the deformation and therefore cannot reproduce the complete three-dimensional stress and strain fields generated by a Berkovich indenter. Since the relaxation parameters ($g_{1}$, $g_{2}$, $\tau_{1}$, and $\tau_{2}$) are sensitive to the local deformation history, these geometric simplifications may slightly influence the identified parameter values. Nevertheless, the equivalent conical indenter with a semi-apex angle of $70.3^\circ$ preserves the same projected contact area as the Berkovich tip, making the axisymmetric formulation a well-established and computationally efficient approximation for nanoindentation analyses. Moreover, the reduced computational cost is particularly advantageous for the iterative inverse optimization procedure adopted in this work, where a full three-dimensional simulation would substantially increase the computational time.

\begin{figure}[H]
    \centering
    \begin{subfigure}[c]{0.45\linewidth}
        \centering
        \includegraphics[width=\linewidth]{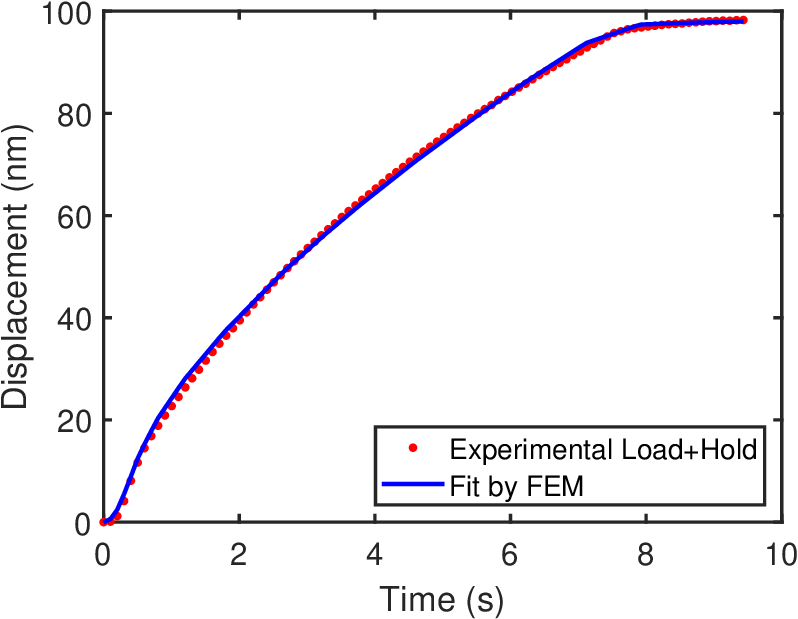}
        \caption{}
        \label{fig10a}
    \end{subfigure}
    \hfill
    \begin{subfigure}[c]{0.45\linewidth}
        \centering
        \includegraphics[width=\linewidth]{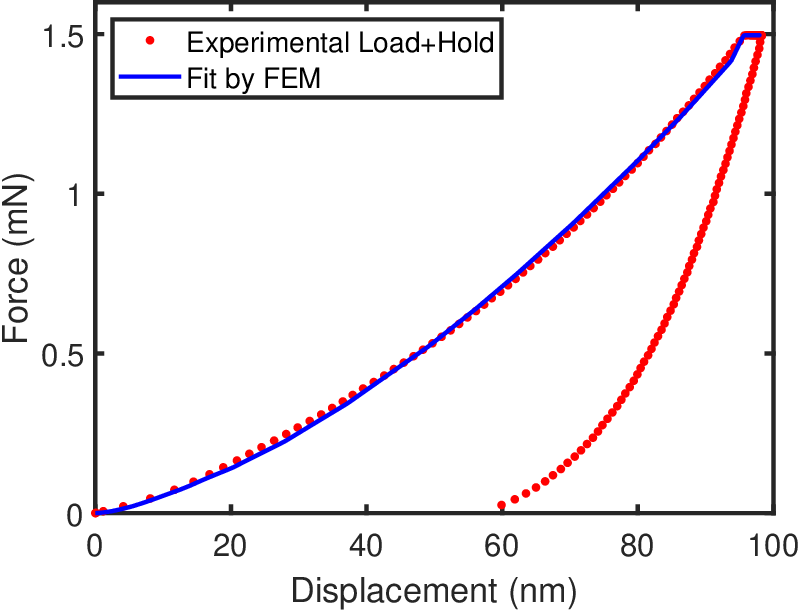}
        \caption{}
        \label{fig10b}
    \end{subfigure}
    \caption{(a) Experimental and FEM-predicted displacement as a function of time using the optimized parameters listed in Table~\ref{tab-params}. (b) Comparison between the experimental and FEM-predicted load-displacement curves.}
    \label{fig10}
\end{figure}

Despite these differences, the high $R^2$ values and low RMSE obtained from both optimizations demonstrate that the analytical and numerical approaches accurately reproduce the experimental force-displacement response. This confirms that different formulations (analytical or numerical) can describe the same macroscopic viscoelastic behavior within their respective modeling frameworks, while yielding different sets of material parameters. Similarly to the previous section, the displacement $h(t)$ predicted using the optimal parameters identified through the finite element inverse optimization was compared with the experimental nanoindentation data, as shown in \hyperref[fig10]{Fig.~\ref{fig10}}.

The displacement-time curve shown in \hyperref[fig10]{Fig.~\ref{fig10a}} demonstrates that the finite element prediction accurately reproduces the experimental response throughout the loading and hold stages, indicating that the identified Prony-series parameters successfully capture the viscoelastic behavior of the coating. Furthermore, the excellent agreement observed in the load-displacement curves (\hyperref[fig10]{Fig.~\ref{fig10b}}) confirms that the identified material parameters provide a reliable representation of the nanoindentation response and validates the proposed inverse finite element identification framework.

Although both the analytical Burgers model and the finite element Prony-series model provide good agreement with the experimental nanoindentation response, they offer complementary advantages. The analytical approach is computationally efficient, converges rapidly, and provides viscoelastic parameters with a clear physical interpretation through the Burgers rheological elements, making it well suited for the rapid identification of constitutive properties. However, it represents the material by an equivalent one-dimensional rheological network and therefore cannot describe the heterogeneous stress and strain fields that develop beneath the indenter. In contrast, the FEM approach explicitly accounts for the indentation contact problem, coating thickness, substrate interaction, and the spatial distribution of stresses and strains, enabling the analysis of local mechanical fields and damage-related phenomena. These additional capabilities are obtained at the expense of a substantially higher computational cost due to the iterative inverse optimization procedure. Therefore, the analytical Burgers model is particularly suitable for the rapid identification of viscoelastic material parameters, whereas the finite element approach provides a more comprehensive description of the local mechanical response and is better suited for investigating complex deformation mechanisms in coated systems.

\subsection{Parametric study on the energy dissipation}
The loading-hold-unload nanoindentation response exhibits a hysteresis loop resulting from the time-dependent viscoelastic behavior of the coating. The enclosed area between the loading and unloading curves represents the mechanical energy dissipated during the indentation cycle and provides valuable information on the contribution of the different viscoelastic mechanisms to the overall material response \cite{100}. Once the optimal viscoelastic parameters had been identified, a parametric sensitivity study was carried out to quantify the influence of each parameter on the hysteresis behavior and the associated energy dissipation. To this end, the complete experimental loading history (loading-hold-unload) was applied in the finite element model using the identified parameter set. Subsequently, a one-parameter-at-a-time sensitivity analysis was performed, in which each parameter was varied individually while the remaining parameters were kept constant, following an approach similar to that reported in \cite{100}.

The sensitivity of the model was quantified by varying one viscoelastic parameter at a time while keeping the remaining parameters fixed at their identified values. For each simulation, the dissipated energy was calculated from the area enclosed by the simulated loading and unloading curves. The ranges considered for each viscoelastic parameter are summarized in Table~\ref{tab-study-params}.

\begin{table}[H]
\centering
\captionsetup{justification=raggedright, singlelinecheck=false}
\caption{Parameters values for the parametric study on the dissipation energy of the viscoelastic model.}
\label{tab-study-params}
\begin{tabular}{lll}
\toprule
\text{Parameters} & \text{Optimal value} & \text{Variation range} \\
\midrule
$v$                     & 0.19                         & / \\
$G_{0}$\ (GPa)          & 52                           & 25, 50, 100 \\
$g_{1}$                 & 0.44                         & 0.22, 0.44, 0.66 \\
& & with ($g_2+g_1\le 1$) \\
$g_{2}/g_{1}$           & 0.27                         & 0.57, 0.91, 1.25 ($g_{2} = 0.25, 0.4, 0.55$) \\
& & with ($g_2+g_1\le 1$) \\
$\tau_{1}$\ (s)         & 0.09                         & 0.01, 0.1, 1 \\
$\tau_{2}/\tau_{1}$     & 294.44 ($\tau_{2} = 26.5$ s) & 56, 278, 556 ($\tau_{2}$ = 5, 25, 50 (s)) \\
\bottomrule
\end{tabular}
\end{table}

\begin{figure}[H]
    \centering
    \includegraphics[width=0.6\linewidth]{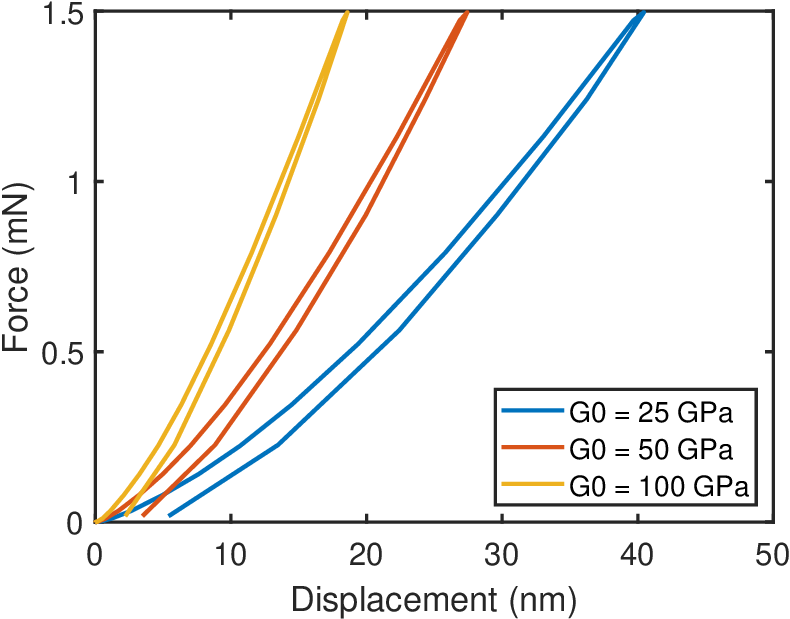}
    \caption{Effect of the initial shear modulus $G_{0}$ on the load-displacement response and energy dissipation during viscoelastic response.}
    \label{fig11}
\end{figure}

As shown in \hyperref[fig11]{Fig.~\ref{fig11}}, under the same maximum applied load, a lower initial shear modulus $G_{0}$ results in a larger indentation displacement. The maximum penetration depths are approximately 40, 27, and 19 nm for $G_{0}$ values of 25, 50, and 100 GPa, respectively. Consequently, the hysteresis loop becomes larger as $G_{0}$ decreases, indicating an increase in the energy dissipated during the indentation cycle. The corresponding dissipated energy, calculated according to $\Delta W = W_{loading}-W_{unloading}$, was used to quantify the influence of the viscoelastic parameters.

The results indicate that the absolute dissipated energy decreases with increasing initial shear modulus. Specifically, $\Delta W$ decreases from 2.55 to 1.65 and 1.10 mN·nm as $G_{0}$ increases from 25 to 50 and 100 GPa, respectively. This behavior is attributed to the higher initial stiffness of the material, which limits the indentation depth and consequently reduces the mechanical work exchanged during the loading-unloading cycle. However, the normalized dissipated energy, expressed as ($100\%\,\frac{\Delta W}{W_{loading}}$), remains nearly constant at approximately $10\% \pm 0.5\%$. This indicates that varying $G_{0}$ mainly scales the overall mechanical response without significantly modifying the relative contribution of viscoelastic dissipation.

\begin{figure}[H]
    \centering
    \begin{subfigure}[c]{0.45\linewidth}
        \centering
        \includegraphics[width=\linewidth]{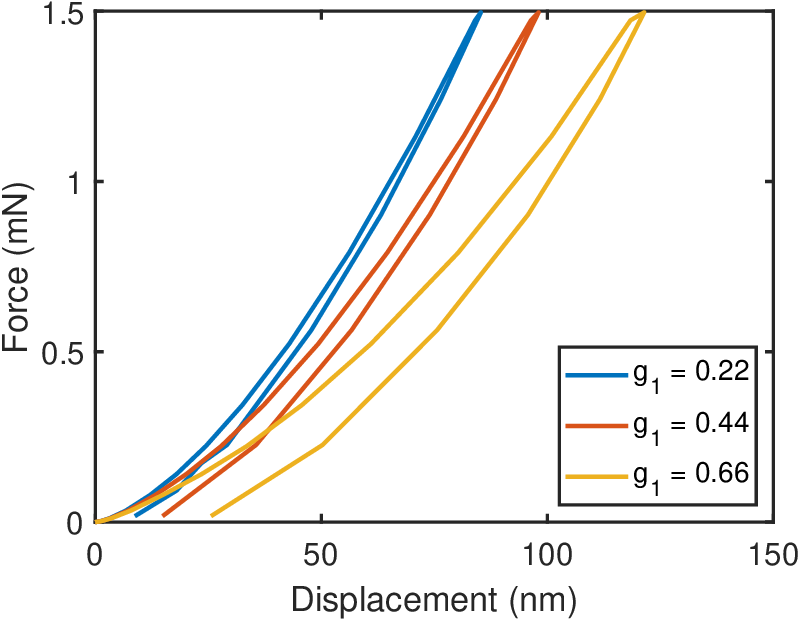}
        \caption{}
        \label{fig12a}
    \end{subfigure}
    \hfill
    \begin{subfigure}[c]{0.45\linewidth}
        \centering
        \includegraphics[width=\linewidth]{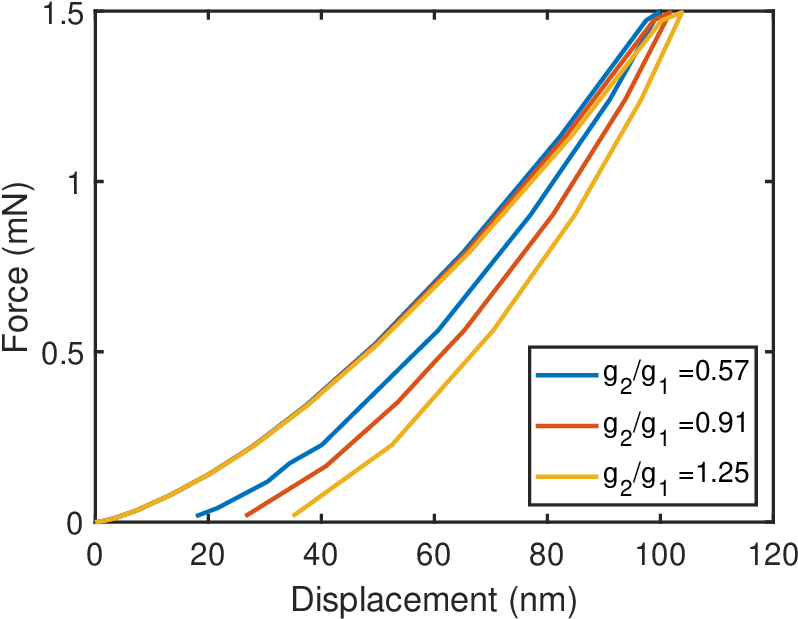}
        \caption{}
        \label{fig12b}
    \end{subfigure}
    \caption{Influence of the relaxation parameters on the viscoelastic load-displacement response: (a) effect of $g_{1}$, (b) effect of the ratio $g_{2}/g_{1}$.}
    \label{fig12}
\end{figure}

Similarly, the force-displacement response was simulated for different values of the shear relaxation coefficient $g_{1}$, while keeping the remaining parameters constant, as shown in \hyperref[fig12a]{Fig.~\ref{fig12a}}. increases the amplitude of the viscoelastic relaxation, leading to a larger reduction of the shear modulus with time according to the Prony-series expression (\hyperref[eq14]{Eq.~\ref{eq14}}). Consequently, the material becomes more compliant during loading and hold, resulting in a greater indentation depth under the same applied load. Increasing $g_{1}$ reduces the shear relaxation modulus described by the Prony-series expression in. As $g_{1}$ increases from 0.22 to 0.44 and 0.66, while satisfying the constraint $g_{1}+g_{2} \le 1$, the maximum indentation depth increases from approximately 85 to 98 and 121 nm, respectively.

The corresponding dissipated energies, calculated from the area enclosed by the loading and unloading curves, are 3.6, 6.5, and 15.0 mN·nm, respectively. Therefore, increasing $g_{1}$ substantially enhances the energy dissipated during the indentation cycle. This behavior is expected because a larger relaxation coefficient increases the fraction of the instantaneous shear modulus that undergoes time-dependent relaxation, thereby producing a larger hysteresis loop. Furthermore, the normalized dissipated energy ($100\%\,\frac{\Delta W}{W_{loading}}$) increases from approximately $7\%$ to $11\%$ and $20\%$, demonstrating that $g_{1}$ has a much stronger influence on viscoelastic dissipation than the initial shear modulus $G_{0}$. This trend is consistent with the observations reported in \cite{100}.

The influence of the ratio between the two Prony-series relaxation coefficients, $g_{2}/g{1}$, on the nanoindentation response was also investigated. As shown in \hyperref[fig12b]{Fig.~\ref{fig12b}}, varying the ratio $g_{2}/g_{1}$ has only a limited effect on the loading stage. The loading curves remain almost identical throughout most of the loading history, with the maximum indentation depth increasing only slightly from approximately 100 to 104 nm as the ratio $g_{2}/g_{1}$ increases from 0.57 to 1.25. This indicates that redistributing the relaxation amplitude between the two Prony branches has only a minor influence on the loading response and, consequently, on the mechanical work applied during loading.

In contrast, the unloading (recovery) stage is strongly influenced by the ratio $g_{2}/g_{1}$. As shown in \hyperref[fig12b]{Fig.~\ref{fig12b}}, increasing $g_{2}/g_{1}$ enlarges the hysteresis loop, indicating greater viscoelastic energy dissipation during the indentation cycle. Consequently, the normalized dissipated energy increases from approximately $18\%$ to $26\%$ and $35\%$ for $g_{2}/g_{1}$ values of 0.57, 0.91, and 1.25, respectively. Therefore, the ratio $g_{2}/g_{1}$ primarily governs the recovery behavior rather than the loading response. This behavior can be interpreted from the Prony-series relaxation function (\hyperref[eq14]{Eq.~\ref{eq14}}), which may be rewritten as: $G(t)=G_{0}-G_{0}g_{1}\left[\left(1-e^{-t/\tau_{1}}\right)+\frac{g_{2}}{g_{1}}\left(1-e^{-t/\tau_{2}}\right)\right]$. For fixed values of $G_{0}$, $g_{1}$, $\tau_{1}$ and $\tau_{2}$, increasing the ratio $g_{2}/g_{1}$ increases the contribution of the second relaxation mechanism, leading to a larger reduction of the relaxation modulus $G(t)$ with time. Consequently, the material remains more compliant during unloading, resulting in a larger hysteresis loop and greater energy dissipation. In the long-time limit ($t\rightarrow \infty$), the relaxation modulus approaches $G(\infty) = G_{0} - G_0\,g_1-\frac{g_{2}}{g_{1}}\,G_0\,g_1$, showing that the equilibrium modulus decreases as the ratio $g_{2}/g_{1}$ increases for fixed $G_{0}$ and $g_{1}$. The equilibrium relaxation modulus decreases linearly with the ratio $g_{2}/g_{1}$ which promotes greater long-term viscoelastic compliance and results in higher energy dissipation.

\begin{figure}[H]
    \centering
    \begin{subfigure}[c]{0.45\linewidth}
        \centering
        \includegraphics[width=\linewidth]{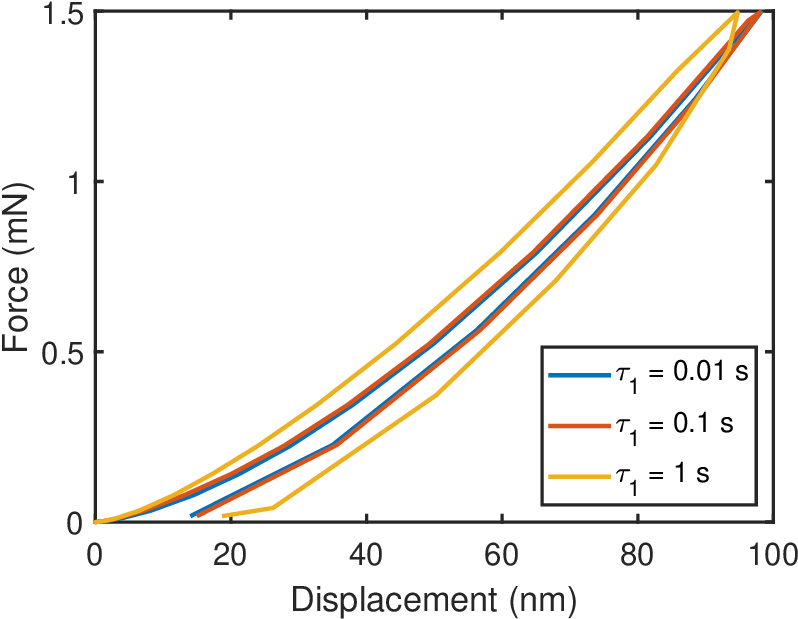}
        \caption{}
        \label{fig13a}
    \end{subfigure}
    \hfill
    \begin{subfigure}[c]{0.45\linewidth}
        \centering
        \includegraphics[width=\linewidth]{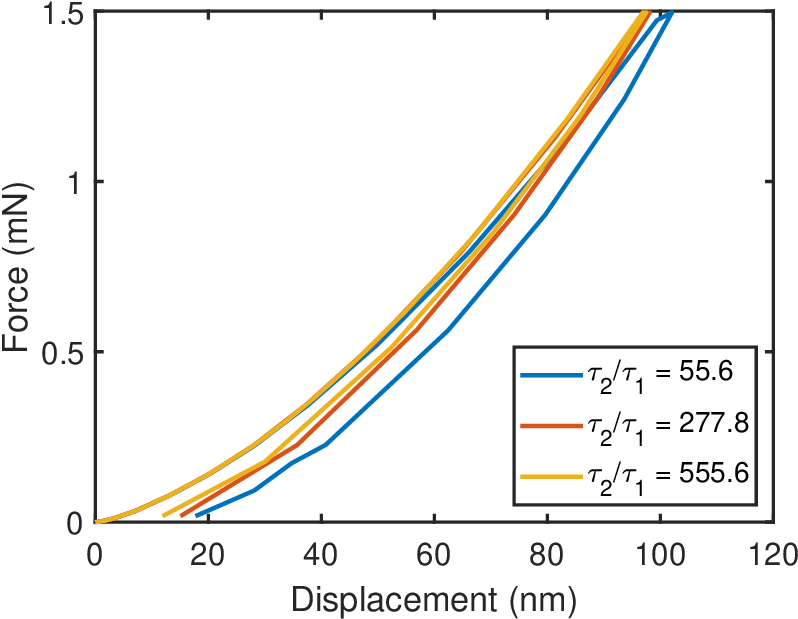}
        \caption{}
        \label{fig13b}
    \end{subfigure}
    \caption{Influence of the relaxation time constants on the viscoelastic load-displacement response and energy dissipation: (a) effect of $\tau_{1}$, (b) effect of the ratio $\tau_{2}/\tau_{1}$.}
    \label{fig13}
\end{figure}

Similarly, the influence of the first relaxation time constant $\tau_{1}$ and the relaxation time ratio $\tau_{2}/\tau_{1}$ on the nanoindentation response was investigated. As shown in \hyperref[fig13a]{Fig.~\ref{fig13a}}, both the loading and unloading stages are affected by the value of $\tau_{1}$. For the two smallest values ($\tau_{1}$ = 0.01 and 0.1 s), the force-displacement curves remain nearly identical, indicating that the first relaxation mechanism occurs much faster than the duration of the indentation experiment and therefore has only a limited influence on the measured response. In contrast, increasing $\tau_{1}$ to 1 s, produces a noticeable shift of both the loading and unloading curves, resulting in a larger hysteresis loop. This behavior indicates that the characteristic relaxation time becomes comparable to the experimental time scale, delaying stress relaxation during loading and elastic recovery during unloading, thereby increasing the energy dissipated during the indentation cycle.

The increase in the hysteresis loop is directly reflected by the energy dissipated during the indentation cycle. The normalized dissipated energy increases from approximately $\%9$ to $11.5\%$ and $29\%$ as $\tau_{1}$ increases from 0.01 s to 0.1 s and 1 s, respectively. These results indicate that the model is only weakly sensitive to small values of $\tau_{1}$, whereas a pronounced increase in energy dissipation occurs when $\tau_{1}$ becomes comparable to the characteristic duration of the nanoindentation experiment. This behavior is consistent with the Prony relaxation function, in which increasing $\tau_{1}$ delays the decay of the exponential relaxation term $e^{-t/\tau_{1}}$. Consequently, stress relaxation and viscoelastic recovery occur over a longer time scale, producing a larger hysteresis loop and increasing the fraction of mechanical work dissipated during the indentation cycle. Therefore, $\tau_{1}$ primarily controls the kinetics of the fast relaxation mechanism and governs the rate of viscoelastic recovery.
 
The influence of the relaxation time ratio $\tau_{2}/\tau_{1}$ on the viscoelastic response was also investigated, as shown in \hyperref[fig13b]{Fig.~\ref{fig13b}}. The force-displacement curves exhibit the greatest variation for relatively small values of the relaxation time ratio. However, as $\tau_{2}/\tau_{1}$ increases, the loading and unloading progressively converge, indicating that the viscoelastic response becomes less sensitive once the second relaxation time is much larger than the first. In this case, the second relaxation mechanism evolves over a time scale much longer than the duration of the nanoindentation experiment and therefore contributes only marginally to the measured response. Consequently, further increases in $\tau_{2}/\tau_{1}$ is several hundred times larger than $\tau_{1}$ produce only minor changes in the apparent stiffness and the hysteresis loop.

The effect of the relaxation time ratio is also reflected in the normalized dissipated energy, which decreases from approximately $20\%$ to $11.5\%$ and $8.5\%$ as $\tau_{2}/\tau_{1}$ increases from 57 to 278 and 556, respectively. This result confirms that increasing the separation between the two relaxation time scales reduces the contribution of the second relaxation mechanism during the nanoindentation cycle, thereby decreasing the energy dissipated through viscoelastic deformation.

This behavior is also consistent with the Prony-series relaxation function (\hyperref[eq14]{Eq.~\ref{eq14}}). Introducing the relaxation time ratio $r_{\tau}=\tau_{2}/\tau_{1}$, the relaxation function can be rewritten as $G(t)=G_{0}\left[1-g_{1}\left(1-e^{-t/\tau_{1}} \right)-g_{2}\left(1-e^{-t/(r_{\tau}\tau_{1})}\right)\right]$. For fixed $G_{0}$, $g_{1}$, $g_{2}$, and $\tau_{1}$, $r_{\tau}$ directly controls the characteristic time of the second relaxation process without affecting its amplitude. A decrease in $r_{\tau}$ shifts the second relaxation process toward shorter times, thereby increasing its contribution over the time scale of the nanoindentation experiment.

\section{Conclusions} \label{conclusion}

This study applies an integrated methodology that combines analytical and finite element (FEM)-based viscoelastic modeling to characterize coating behavior under nanoindentation. The analytical approach, based on the Burgers model, provides a direct correlation between the viscoelastic response and the indentation load–displacement data, allowing the extraction of rheological parameters describing the time-dependent deformation. Complementarily, the FEM implementation in ABAQUS using a 2D axisymmetric model and Prony-series relaxation formulation 
enabled the simulation of indentation behavior under identical loading conditions. An automated inverse optimization routine using Nelder–Mead 
algorithm minimized the discrepancies (quadratic error) between experimental and simulated responses.

The results demonstrated an agreement between experimental and analytical data, as well as between experimental and numerical simulations confirming the robustness of the inverse optimization approach. Although some discrepancies were observed between the analytical and FEM-derived viscoelastic parameters, both approaches consistently reproduced the experimental load-displacement behavior and captured the time-dependent response of the coating. The simplified axisymmetric FEM model efficiently reproduced the experimental results while considerably reducing the computational cost, making it a practical alternative to full three-dimensional simulations for iterative parameter identification. In addition, the numerical framework provides access to local stress and strain distributions that are inaccessible through analytical formulation alone, making the framework suitable for future studies of coating damage and failure.

A complementary parametric study was performed to investigate the influence of the Prony-series parameters on the viscoelastic response and energy dissipation during the indentation cycle. The results demonstrated that the initial shear modulus and the relaxation coefficients primarily control the material compliance and the hysteresis loop, whereas the relaxation time constants govern the kinetics of viscoelastic recovery. These findings provide a better physical understanding of the contribution of each rheological parameter to the indentation response and constitute useful guidelines for the identification and optimization of viscoelastic material properties.

Overall, the proposed integrated framework provides an efficient and reliable methodology for identifying viscoelastic properties from nanoindentation experiments while offering a computationally efficient alternative for inverse parameter identification. Although demonstrated for WSe\textsubscript{2} coatings, the methodology is readily applicable to a broad range of viscoelastic coatings and polymeric materials, provided that their mechanical behavior can be described by an appropriate viscoelastic constitutive model. This work also establishes a solid foundation for future studies involving temperature-dependent viscoelasticity, damage evolution, and tribological performance of advanced coating systems.

\appendix

\setcounter{figure}{0}
\renewcommand{\thefigure}{A.\arabic{figure}}

\section{Simplex Scheme} \label{Appendix A}

The simplex can shift its position while approximately maintaining its size or shrinking as it approaches an optimum. Let $x_h$  represents the 
vertex with the highest function value, $x_s$ the vertex with the second highest function value, and $x_l$ the vertex with the lowest function 
value. Let $\bar{x}$ denotes the mean of all vertices except the highest point $x_h$. For any design point $x_\theta$, let $y_\theta = 
f(x_\theta)$ be its corresponding objective function value. A single iteration of the method could evaluate four simplex operations.

\noindent\textbf{Reflection:} The reflection point $x_r$ is computed as $x_r = \bar{x} + \alpha(\bar{x}-x_h )$, which reflects the vertex with 
the highest function value over the centroid of the remaining vertices. This operation usually moves the simplex away from higher-value regions 
toward potentially lower-value regions. The reflection coefficient $\alpha$ is positive, typically set to $\alpha = 1$.

\noindent\textbf{Expansion:} The expansion point $x_e$ is calculated as $x_e = \bar{x} + \beta(x_r-\bar{x})$, like reflection, but it extends 
the reflected point even further. This operation is performed when the reflected point has a lower objective function value than all other 
points in the simplex, indicating a promising direction. The expansion coefficient $\beta$ is greater than $\max(1,\alpha)$, and is typically 
set to $\beta = 2$.

\noindent\textbf{Contraction:} The contraction point $x_c$ is computed as $x_c = x + \gamma(x_h - \bar{x})$, where the simplex shrinks by 
moving closer to the centroid and away from the worst point $x_h$. This operation is used when neither reflection nor expansion provides a 
better point. The contraction coefficient $\gamma$ lies within the range (0,1) and is typically set to $\gamma = 0.5$.

\noindent\textbf{Shrinkage:} In the shrinkage operation, all points in the simplex are moved closer to the best point $x_l$. This typically 
involves halving the separation distance between each point and $x_l$. Shrinkage is applied when other operations (reflection, expansion, and 
contraction) fail to produce a better result, helping the simplex focus on a smaller region near the best point.

\begin{figure}[H]
    \centering
    \includegraphics[width=\linewidth]{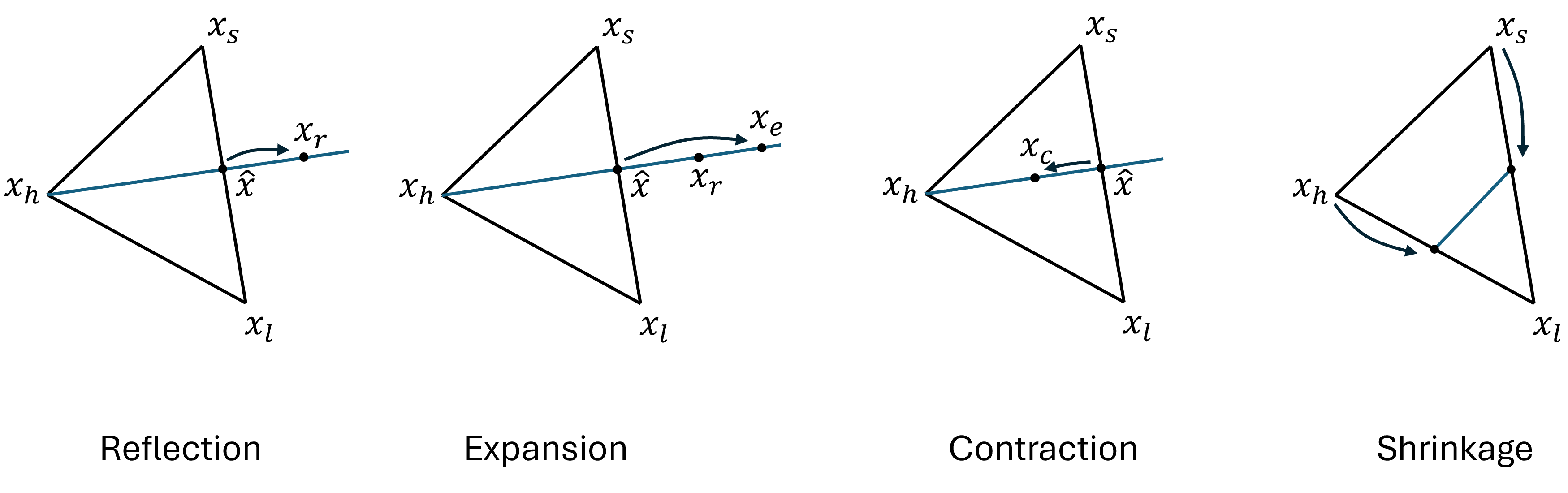}
    \caption{Algorithm actions of reflection, expansion, contraction and shrinkage of simplex over iterations illustrated in \cite{30}.}
    \label{fig14}
\end{figure}

\section*{CRediT authorship contribution statement}

\textbf{Mohamed Bensalem:} Writing – review \& editing, Writing – original draft, Conceptualization, Methodology, Simulations, Investigation, Algorithms, Formal analysis.
\textbf{Fateh Bahadur:} Review \& editing, Conceptualization.
\textbf{Yue Wang:} Preparing sample, Review \& editing, Conceptualization.
\textbf{Nabil Daghbouj:} Writing – review \& editing, Investigation, Methodology, Formal analysis, Supervision.
\textbf{Tomas Polcar:} Writing – review \& editing, Investigation, Supervision, Funding acquisition. 

\noindent {\bf{Declaration of competing interest}}

The authors declare that they have no known competing financial interests or personal relationships that could have appeared to influence the 
work reported in this paper.

\noindent {\bf{Acknowledgement}}

The project was supported by the Czech Science Foundation (project No. 23-07785S); the computing part was partially co-funded by the European 
Union under the project Robotics and advanced industrial production (reg. no. CZ.02.01.01/00/22\textunderscore008/0004590). This work was 
supported by the Ministry of Education, Youth and Sports of the Czech Republic through the e-INFRA CZ (ID:90254). CzechNanoLab project 
LM2023051 funded by MEYS CR is gratefully acknowledged for the financial support of the measurements/sample fabrication at CEITEC Nano Research 
Infrastructure.

\noindent {\bf{Data availability}}

Data will be made available on request.

\bibliographystyle{elsarticle-num}

\bibliography{References}

\end{document}